\documentclass[twocolumn]{aastex631}
\usepackage{graphicx}

\usepackage[utf8]{inputenc}

\usepackage{amsmath}	

\newcommand{\swift}{\textit{Swift~}}

\begin{document}

\title{Swift-BAT GUANO follow-up of gravitational-wave triggers in the third  LIGO--Virgo--KAGRA observing run}


\author[0000-0003-0852-3685]{Gayathri Raman}
\affiliation{Department of Astronomy and Astrophysics, 
The Pennsylvania State University, 525 Davey Lab, University Park, PA 16802, USA}

\author[0000-0003-0020-687X]{Samuele Ronchini}
\affiliation{Department of Astronomy and Astrophysics, 
The Pennsylvania State University, 525 Davey Lab, University Park, PA 16802, USA}

\author[0000-0001-5229-1995]{James Delaunay}
\affiliation{Department of Physics and Astronomy, University of Alabama, Tuscaloosa, AL 35487, USA}
\affiliation{Department of Astronomy and Astrophysics, 
The Pennsylvania State University, 525 Davey Lab, University Park, PA 16802, USA}

\author[0000-0002-2810-8764]{Aaron Tohuvavohu}
\affiliation{Department of Astronomy \& Astrophysics, University of Toronto, Toronto, ON M5S 3H4}
\affiliation{Dunlap Institute for Astronomy \& Astrophysics, University of Toronto, Toronto, ON M5S 3H4}

\author[0000-0002-6745-4790]{Jamie A. Kennea}
\affiliation{Department of Astronomy and Astrophysics, 
The Pennsylvania State University, 525 Davey Lab, University Park, PA 16802, USA}

\author[0000-0002-4299-2517]{Tyler Parsotan}
\affiliation{Astrophysics Science Division, NASA Goddard Space Flight Center, Greenbelt, MD 20771, USA}


\author{Elena Ambrosi}
\affiliation{INAF – IASF-Palermo, via Ugo La Malfa 153, 90146 Palermo PA, Italy}

\author{Maria Grazia Bernardini}
\affiliation{INAF-Osservatorio Astronomico di Brera, Via E. Bianchi 46, 23807 Merate, LC, Italy}

\author{Sergio Campana}
\affiliation{INAF-Osservatorio Astronomico di Brera, Via E. Bianchi 46, 23807 Merate, LC, Italy}

\author{Giancarlo Cusumano}
\affiliation{INAF – IASF-Palermo, via Ugo La Malfa 153, 90146 Palermo PA, Italy}

\author[0000-0002-5042-1036]{Antonino D'A\`{\i}}
\affiliation{INAF – IASF-Palermo, via Ugo La Malfa 153, 90146 Palermo PA, Italy}

\author{Paolo D'Avanzo}
\affiliation{INAF-Osservatorio Astronomico di Brera, Via E. Bianchi 46, 23807 Merate, LC, Italy}

\author{Valerio D’Elia}
\affiliation{Space Science Data Center (SSDC) – Agenzia Spaziale Italiana
(ASI), 00133 Roma, Italy}
\affiliation{INAF – Osservatorio Astronomico di Roma, Via Frascati 33, 00040 Monte Porzio Catone, Italy}

\author{Massimiliano De Pasquale}
\affiliation{MIFT Department, Polo Papardo, University of Messina, Viale Ferdinando Stagno d'Alcontres, 31, 98166 Messina, Italy}

\author{Simone Dichiara}
\affiliation{Department of Astronomy and Astrophysics, 
The Pennsylvania State University, 525 Davey Lab, University Park, PA 16802, USA}

\author[0000-0002-8465-3353]{Phil Evans}
\affiliation{School of Physics and Astronomy, University of Leicester, University Road, Leicester LE1 7RH, UK}

\author{Dieter Hartmann}
\affiliation{Department of Physics \& Astronomy, Clemson University, Kinard Lab of Physics, Clemson, SC 29634, USA}

\author{Paul Kuin}
\affiliation{Mullard Space Science Laboratory, University College London, Holmbury St. Mary, Dorking, Surrey RH5 6NT, UK}

\author{Andrea Melandri}
\affiliation{INAF-Osservatorio Astronomico di Roma, Via di Frascati 33, 00040 Monte Porzio Catone, RM, Italy}

\author{Paul O'Brien}
\affiliation{School of Physics and Astronomy, University of Leicester, University Road, Leicester LE1 7RH, UK}

\author{Julian P. Osborne}
\affiliation{School of Physics and Astronomy, University of Leicester, University Road, Leicester LE1 7RH, UK}

\author{Kim Page}
\affiliation{School of Physics and Astronomy, University of Leicester, University Road, Leicester LE1 7RH, UK}

\author[0000-0001-7128-0802]{David M. Palmer}
\affiliation{Los Alamos National Laboratory, PO Box 1663, Los Alamos New Mexico 87545}

\author{Boris Sbarufatti}
\affiliation{INAF - Osservatorio Astronomico di Brera, Via E. Bianchi 46, 23807 Merate, Italy}

\author{Gianpiero Tagliaferri}
\affiliation{INAF-Osservatorio Astronomico di Brera, Via E. Bianchi 46, 23807 Merate, LC, Italy}

\author[0000-0002-1869-7817]{Eleonora Troja}
\affiliation{University of Rome Tor Vergata, via Cracovia 50, 00100 Roma, Italy}

\collaboration{3000}{The LIGO Scientific Collaboration, the Virgo Collaboration, and the KAGRA Collaboration}

\author{A.~G.~Abac}
\affiliation{Max Planck Institute for Gravitational Physics (Albert Einstein Institute), D-14476 Potsdam, Germany}
\author{R.~Abbott}
\affiliation{LIGO Laboratory, California Institute of Technology, Pasadena, CA 91125, USA}
\author{H.~Abe}
\affiliation{Graduate School of Science, Tokyo Institute of Technology, 2-12-1 Ookayama, Meguro-ku, Tokyo 152-8551, Japan}
\author{I.~Abouelfettouh}
\affiliation{LIGO Hanford Observatory, Richland, WA 99352, USA}
\author{F.~Acernese}
\affiliation{Dipartimento di Farmacia, Universit\`a di Salerno, I-84084 Fisciano, Salerno, Italy}
\affiliation{INFN, Sezione di Napoli, I-80126 Napoli, Italy}
\author[0000-0002-8648-0767]{K.~Ackley}
\affiliation{University of Warwick, Coventry CV4 7AL, United Kingdom}
\author[0000-0001-5525-6255]{C.~Adamcewicz}
\affiliation{OzGrav, School of Physics \& Astronomy, Monash University, Clayton 3800, Victoria, Australia}
\author{S.~Adhicary}
\affiliation{The Pennsylvania State University, University Park, PA 16802, USA}
\author[0000-0002-4559-8427]{N.~Adhikari}
\affiliation{University of Wisconsin-Milwaukee, Milwaukee, WI 53201, USA}
\author[0000-0002-5731-5076]{R.~X.~Adhikari}
\affiliation{LIGO Laboratory, California Institute of Technology, Pasadena, CA 91125, USA}
% ...

\begin{abstract}
We present results from a search for X-ray/gamma-ray counterparts of gravitational-wave (GW) candidates from the third observing run (O3) of the LIGO--Virgo--KAGRA (LVK) network using the \textit{Swift} Burst Alert Telescope (\textit{Swift}-BAT). The search includes 636 GW candidates received in low latency, 86 of which have been confirmed by the offline analysis and included in the third cumulative Gravitational-Wave Transient Catalogs (GWTC-3). Targeted searches were carried out on the entire GW sample using the maximum--likelihood NITRATES pipeline on the BAT data made available via the GUANO infrastructure. We do not detect any significant electromagnetic emission that is temporally and spatially coincident with any of the GW candidates. We report flux upper limits in the $15$--$350$~keV band as a function of sky position for all the catalog candidates. For GW candidates where the \textit{Swift}-BAT false alarm rate is less than $10^{-3}$ Hz, we compute the GW--BAT joint false alarm rate. Finally, the derived \textit{Swift}-BAT upper limits are used to infer constraints on the putative electromagnetic emission associated with binary black hole mergers.
\end{abstract}

\section{Introduction}

The discovery of gravitational waves (GWs) from coalescing binary black holes (BBH) by the Laser Interferometer Gravitational-Wave Observatory (LIGO) opened a new window to the Universe \citep{LIGO2016PhRvXBBHO1}. In addition to GWs, compact binary mergers with at least one neutron star (NS) component are likely to generate electromagnetic (EM) radiation (e.g., \citealt{nak2020PhR...886....1N, Kyutoku2021LRR}). Coincident detection of EM emission from compact binary mergers provides a complete picture of the merger process and can have huge implications for our understanding of the Universe.
Such coincidences play a crucial role in tracing the properties of the source host galaxy \citep{Troja2017Natur,Alexander2017ApJ}, mitigating degeneracies in GW parameter estimation \citep{Abbott2017PhRvL119p1101A,Hughes2003CQGra,Wang2021ApJ}, placing constraints on the NS equation of state \citep{Bauswein2017ApJ,Radice2018ApJ}, and investigating the expansion rate of the Universe, thereby testing cosmological models \citep{Abbott2017PhRvL119p1101A,2017Natur551standardsiren,Kenta2019NatAs,Nissanke2013arXiv,Schutz1986Natur.323..310S}. Additionally, they allow for the measurement of arrival time differences between photons and gravitons, providing limits to the mass of a graviton, exploring potential violations to the equivalence principle and Lorentz invariance \citep{LIGO2017ApJspeedofgrav}.

The joint detection of the first GW event consistent with a binary NS (BNS) coalescence GW170817 \citep{Abbott2017PhRvL119p1101A}, and a coincident short gamma-ray burst GRB 170817A \citep{2017ApJ...848L..14G,Savchenko2017ApJ}, accompanied by the optical/infrared \textit{kilonova} counterpart AT 2017gfo  \citep{Arcavi2017Natur,Coulter2017Sci,Tanvir2017ApJ,Evans2017SciGw170817, Pian2017Natur, Smartt2017Nature,Drout2017Sci,Cowperthwaite2017ApJ} and the GRB afterglow (in the X-rays: \citealt{Troja2017Natur,Margutti2017ApJ} and radio: \citealt{HallinanCorsi2017Sci...358.1579H}), together ushered in a new era in the field of multi-messenger astrophysics and forever impacted our comprehension of compact binary coalescences (CBC) involving an EM counterpart. Massive coordinated EM follow-up efforts were dedicated to deeply monitor the error regions derived from the joint sky localization of GW detectors and high-energy satellites, helping to reduce the initial three detector network sky localization from 28 deg$^2$ to within a few arcseconds of the host galaxy NGC 4993 \citep{Abbott2017bApJ,Abbott2017PhRvL119p1101A}. The spectacular spectral and light curve evolution of this transient \citep{Abbott2017bApJ,Villar2017ApJ} suggested that this explosive event was an active site for r-process nucleosynthesis \citep{Pian2017Natur,Smartt2017Nature,Coulter2017Sci,Drout2017Sci} (for a detailed review of the multimessenger observations of GW170817, see, e.g., \citealt{Nakar2020PhR,Margutti2021ARAA}).

The expected EM counterpart emission from BNS or neutron star--black hole (NSBH) mergers can potentially be weak due to various factors such as considerable source distances, an off-axis viewing angle, or limited amount of ejected mass. For the specific case of GW170817, despite a coincident GRB detection, it took nearly half a day to localize the host galaxy and begin observations of the kilonova \citep{Abbott2017PhRvL119p1101A}. Prompt targeted searches around the GW trigger times, leveraging facilities with enhanced localization capabilities, can refine search strategies and assist optical or infrared (IR) facilities in correctly identifying and pursuing transient candidates for subsequent follow-up studies. In addition to prompt searches, \textit{Fermi}-GBM analysis of triggers from the first and second LIGO--Virgo observing runs showed that targeted \textit{offline} searches are capable of recovering additional candidate joint events that may be of astrophysical relevance \citep{Hamburg2020ApJ...893..100H, Marion2023arXiv230604373P}. Temporal and spatial coincidence information can be used to derive the joint false alarm rate (FAR). These estimates have the potential to elevate subthreshold triggers in either the GW or GRB domains to the status of an above-threshold candidate detection \citep{2019ApJ...876L...4N}.

Unlike \textit{Fermi}, \swift has been for a long time incapable of relaying a continuous stream of event mode data to the ground in real time. Such a capability was enabled through GUANO (Gamma-ray Urgent Archiver for Novel Opportunities, described in Section \ref{swift_sec}) \citep{GUANOTohuvavohu2020}, which recovers event data from the \swift Burst Alert Telescope (BAT, \citealt{Barthelmy2005}), that then get processed by the Non-Imaging Transient Reconstruction And TEmporal Search (NITRATES,  \citealt{NITRATES2022}) pipeline (see Section \ref{sec:target}) to search for subthreshold transient candidates.\footnote{Live reporting of the status of the real time \textit{Swift}-BAT subthreshold analysis can be found at \url{https://guano.swift.psu.edu}, where the user can monitor all the triggers ingested by GUANO and visualize the main results of the NITRATES analysis.} In addition to other astronomical transients, such as GRBs, fast radio bursts (FRBs), and high-energy neutrinos, the GUANO-NITRATES infrastructure performs targeted searches on GW events communicated by the LVK Collaboration, to detect possible GRBs associated with CBCs.

The impact and potential of \textit{Swift}-BAT subthreshold searches are crucial for multi-messenger related goals. Indeed, deeper targeted searches increase the joint detection horizon, thus enhancing the probability of finding weak EM counterparts of CBCs in the hard X-ray domain. Moreover, thanks to the high spatial accuracy enabled by the BAT coded mask, subthreshold searches open the possibility of recovering the position of the candidate EM event at the precision level of a few arcminutes, fundamental to drive the subsequent follow-up with ground and space-based EM facilities. 

Currently, the targeted search analysis carried out thanks to GUANO, has enabled the discovery of more than 35 GRBs with arcminute localization. A total of 7 of the detected GRBs have a duration $<2$~s (e.g., \citealt{2020GCN27444short,2022GCN32167short}), hence they are potentially associated with CBCs containing at least one NS. GUANO data have also been used for the localization of 29 long GRBs through imaging (e.g., \citealt{2021GCN30130longarcminslew}) and non-imaging analysis techniques (e.g., \citealt{2021GCN30302longarcmin,2022GCN32375longarcmin}). GRB 220107A, detected during BAT slew and localized with arcminute precision, enabled the first optical redshift measured using GUANO data \citep{2022GCN31402arcminoptz}.  The arcminute localization of GRB 211106A enabled prompt multiband follow up and led to the discovery of the first afterglow in the millimeter band from a short GRB \citep{2021GCN31049arcminshortmmz}.  With GUANO, one can additionally recover coarse localization information on GRB-like transients that originate from outside the BAT field of view (FOV; e.g., \citealt{2023GCN34747shortofov}). 

In addition to the application to real-time analysis, the availability of BAT data enables us to perform a systematic, deeper targeted search focused on archival LVK triggers. The goal of this study is to perform such an analysis on all the LVK triggers received during the third LIGO--Virgo observing run, during which the GUANO pipeline started to be fully operational.
The run duration was comprised of two segments: O3a, which operated from April 1, 2019, 15:00 UTC to October 1, 2019, 15:00, and O3b which operated from November 1, 2019, 15:00 UTC, to March 27, 2020, 17:00 UTC. The alerts distributed during O3 were reporting the following parameters: FAR, the signal classification (CBC or unmodeled Burst), and the associated astrophysical probabilities. The results of O3 are summarized in Gravitational-Wave Transient Catalog data releases GWTC-2 \citep{GWTC2-2021PhRvX}, GWTC-2.1 \citep{GWTC-2.1PhysRevD.109.022001}, and GWTC-3 \citep{GWTC3-2023PhRvX}.

In this work, we use \textit{Swift}-BAT observations to carry out offline targeted subthreshold searches for EM counterparts of the GW triggers obtained during O3. The rest of the paper is organized as follows: In Section \ref{swift_sec}, we describe the \textit{Swift}-BAT instrument and its new capabilities, and in Section \ref{gw_sam}, we provide details about the GW trigger sample used for the analysis. In Section \ref{sec:target} we summarize the targeted search method adopted for the analysis. We present the results from our targeted search analysis on the various subcategories of triggers in Section \ref{nit_res}, and discuss the scientific interpretation in Section \ref{science}.  

\section{\textit{S\MakeLowercase{wift}}-BAT} 
\label{swift_sec}
The Neil Gehrels \textit{Swift} Observatory (henceforth, \textit{Swift}) is a GRB-focused mission launched in 2004, with three onboard payloads -- the Burst Alert Telescope (BAT), the X-ray Telescope (XRT), and the UltraViolet/Optical Telescope (UVOT) -- covering the EM spectrum from hard X-rays and gamma-rays all the way to the optical \citep{Gehrels2004}. The BAT instrument \citep{Barthelmy2005} is a hard X-ray coded mask imager with a wide FOV that operates in the broad energy band of 15--350~keV. It is the primary instrument that detects GRBs and performs an onboard imaging analysis via a cross-correlation between the spatial pattern of the counts across the detector array and the pattern of the coded mask. The sensitivity of BAT is capable of providing arcminute localizations of GRBs triggered onboard \citep{Gehrels2004}. Due to the lack of continuous downlinking of timing, spatial, and energy information for each detector count (event mode data), carrying out targeted searches offline has not been possible in the past. A new infrastructure, called GUANO, was incorporated into the \textit{Swift}-BAT operations in 2019. Details of the GUANO operations can be found in \citet{GUANOTohuvavohu2020}. From the outset, GUANO has demonstrated that recovering event mode data from astrophysically compelling time windows can enhance the overall transient detection rate and sensitivity of BAT \citep{GUANOTohuvavohu2020}.

\section{Gravitational-Wave Trigger Sample}
\label{gw_sam}
This paper focuses on the \textit{Swift}-BAT subthreshold analysis of a sample of GW triggers with a FAR$<$2 per day, distributed by the LVK Collaboration during O3. The subthreshold GW alerts were received by the EM follow-up groups that were part of a Memorandum of Understanding with the LVK Collaboration. For candidates found with CBC search pipelines \citep{2021ApJ...923..254D,Surabhi2019arXiv, Messick2017PhRvD, Aubin2021CQGra, Nitz2018PhRvD, Hooper2012PhRvD} and Burst search pipelines \citep{Klimenko2005PhRvD,Klimenko2016PhRvD}, the alerts contain basic information about the GW FAR, the probability of the candidate being astrophysical ($p_{\rm astro}$) and trigger time. In the case of CBC candidates, the alerts received in low latency report the $p_{\rm astro}$ split in the four CBC classes: BBH, BNS, NSBH, and Mass Gap. The Mass Gap category includes CBC candidates in which at least one component has a mass in the range [3--5] $M_{\odot}$. Using the GW trigger information received in low latency, we can further perform a search for associations in BAT data.

From the list of 1552 alerts that were communicated via low latency channels, we obtained successful GUANO data dumps for 636 triggers. The GW information of the candidates received in low latency are reported in Table ~\ref{low-latency}. The FAR and the $p_{\rm astro}$ classification reported here correspond to the preferred event, namely the one with the highest SNR. \textit{Swift}-BAT event mode data coincident with the GW trigger time were made available for these triggers in real time via the GUANO data dumps. Post-processing on the data was carried out on this sample from O3 using the NITRATES analysis pipeline (see Section \ref{sec:target}).

Out of the 636 low-latency alerts, a total of 86 GW candidates have been confirmed by the offline analysis and included in the GWTC-2.1 and GWTC-3 data releases \citep{GWTC-2.1PhysRevD.109.022001, GWTC3-2023PhRvX}. Among the 86 confirmed candidates, 14 triggers have $p_{\rm astro}>0.5$ and 72 triggers have $p_{\rm astro}<0.5$. We indicate the details of the confirmed candidates with $p_{\rm astro}>0.5$ and $p_{\rm astro}<0.5$ in Table \ref{tab:gwtc3} and Table \ref{tab:confirmed-subthresh}, respectively. The values of FAR, $p_{\rm astro}$ and CBC Class given in Table \ref{tab:gwtc3} and Table \ref{tab:confirmed-subthresh} are derived from offline analyses, as reported in GWTC-2.1 and GWTC-3 data releases, hence are considered more reliable than the values obtained in low latency, reported in Table \ref{low-latency}. The FAR and the $p_{\rm astro}$ classification reported in Tables~\ref{tab:gwtc3} and \ref{tab:confirmed-subthresh} come from the pipeline that gives the highest value of $p_{\rm astro}$. For high significance events, if multiple pipelines derive a $p_{\rm astro}\simeq1$, we select the one with highest SNR. According to these rules, in the case of S200225q, reported in Table~\ref{tab:gwtc3}, the selected pipeline is cWB, but the event is classified as CBC. The CBC class is determined by the highest among $p_{\rm BBH}$, $p_{\rm NSBH}$ and $p_{\rm BNS}$. In Tables \ref{tab:gwtc3} and \ref{tab:confirmed-subthresh} we report the value of $p_{\rm Class}$ defined as $\max[p_{\rm BBH}, p_{\rm NSBH}, p_{\rm BNS}]$.
It is possible that some candidates marked as ``Burst" in Table ~\ref{low-latency} are then re-classified as ``CBC" by the offline analysis. Therefore, for each catalog event the most updated group is the one reported in Tables \ref{tab:gwtc3} and \ref{tab:confirmed-subthresh}.

\begin{figure*}
    \centering
    \includegraphics[width=0.5\columnwidth]{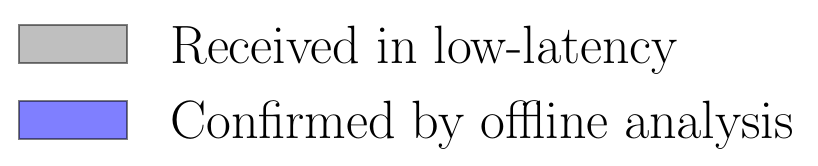}\\
    \includegraphics[width=0.9\columnwidth]{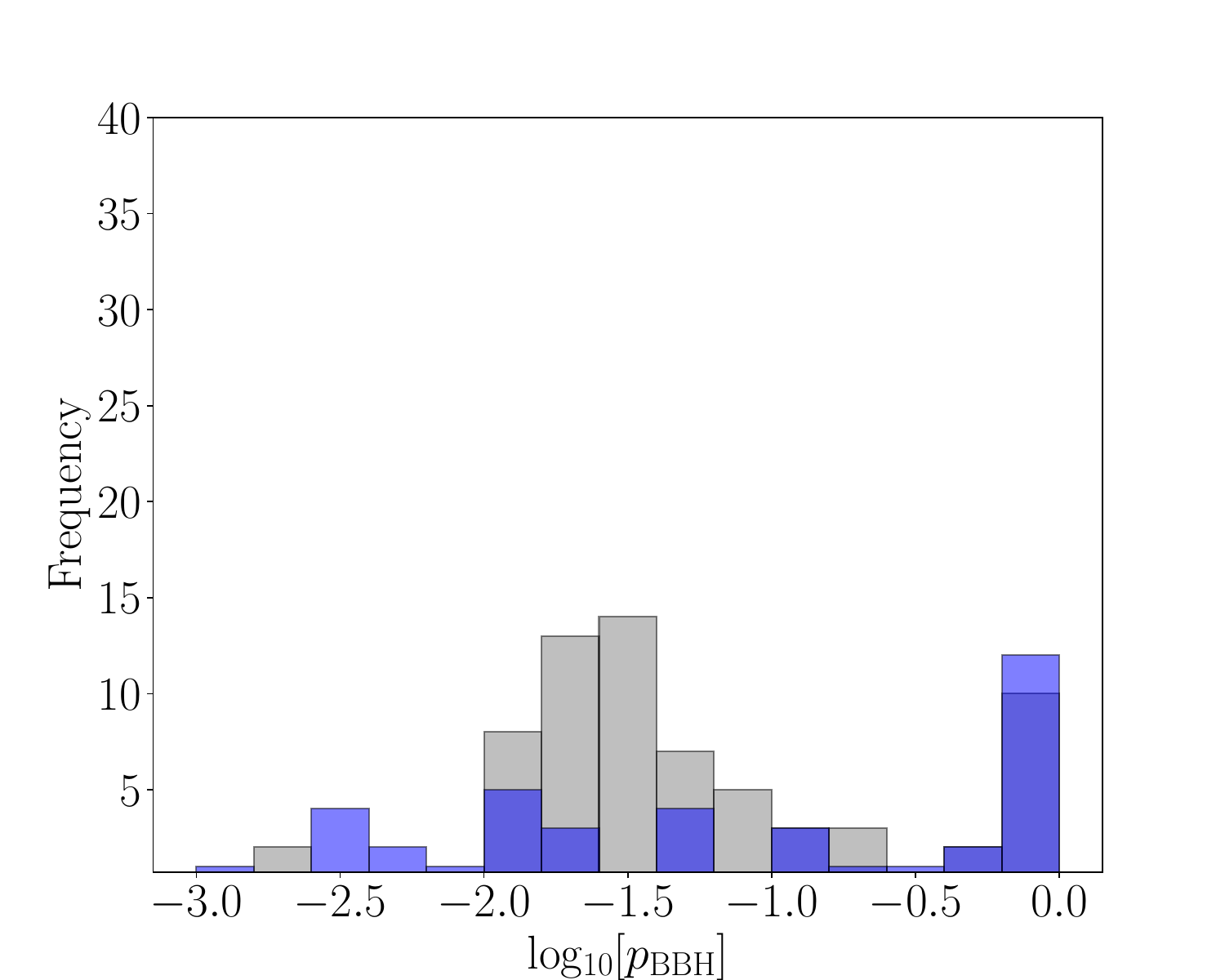}
    \includegraphics[width=0.9\columnwidth]{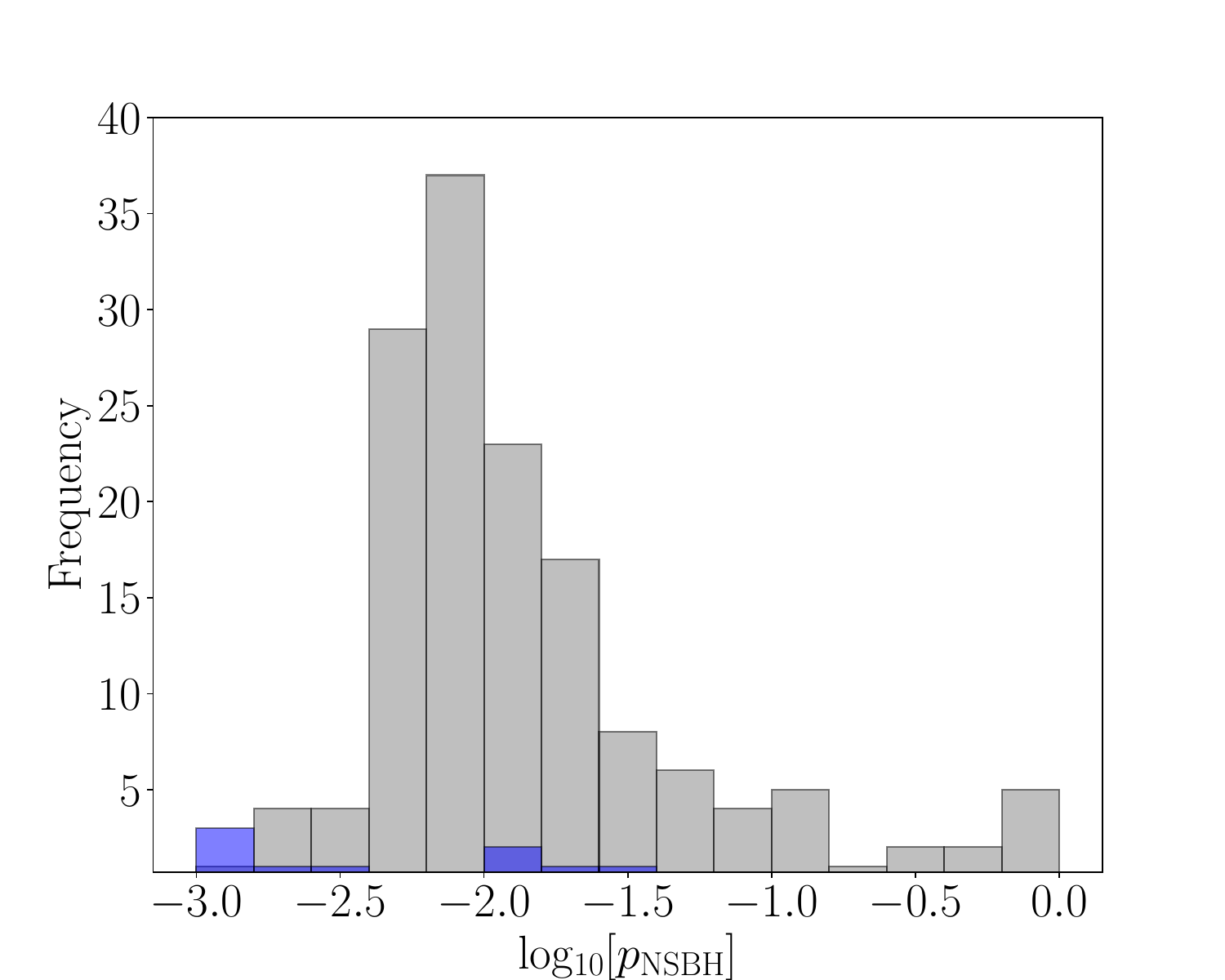}

    \includegraphics[width=0.9\columnwidth]{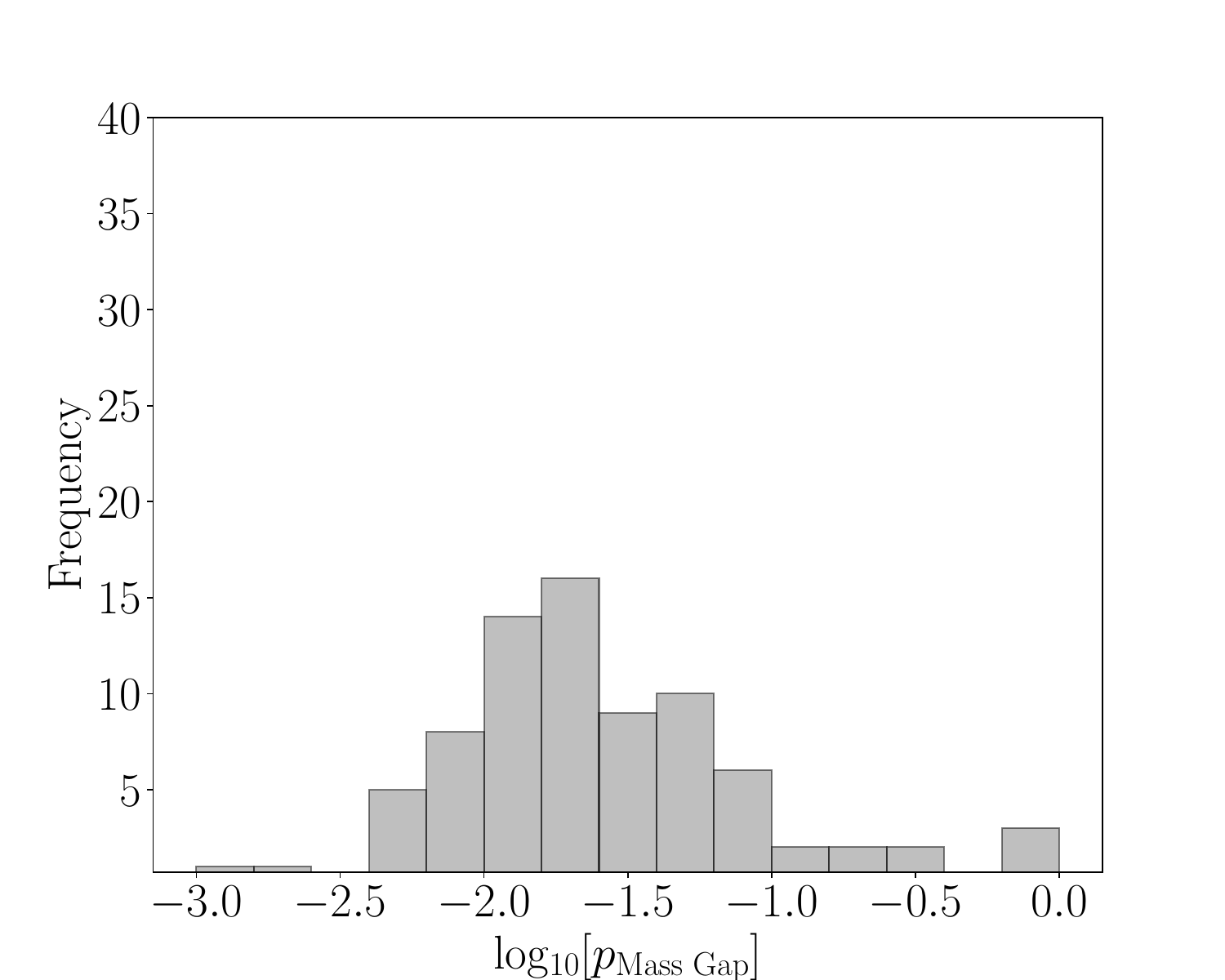}
    \includegraphics[width=0.9\columnwidth]{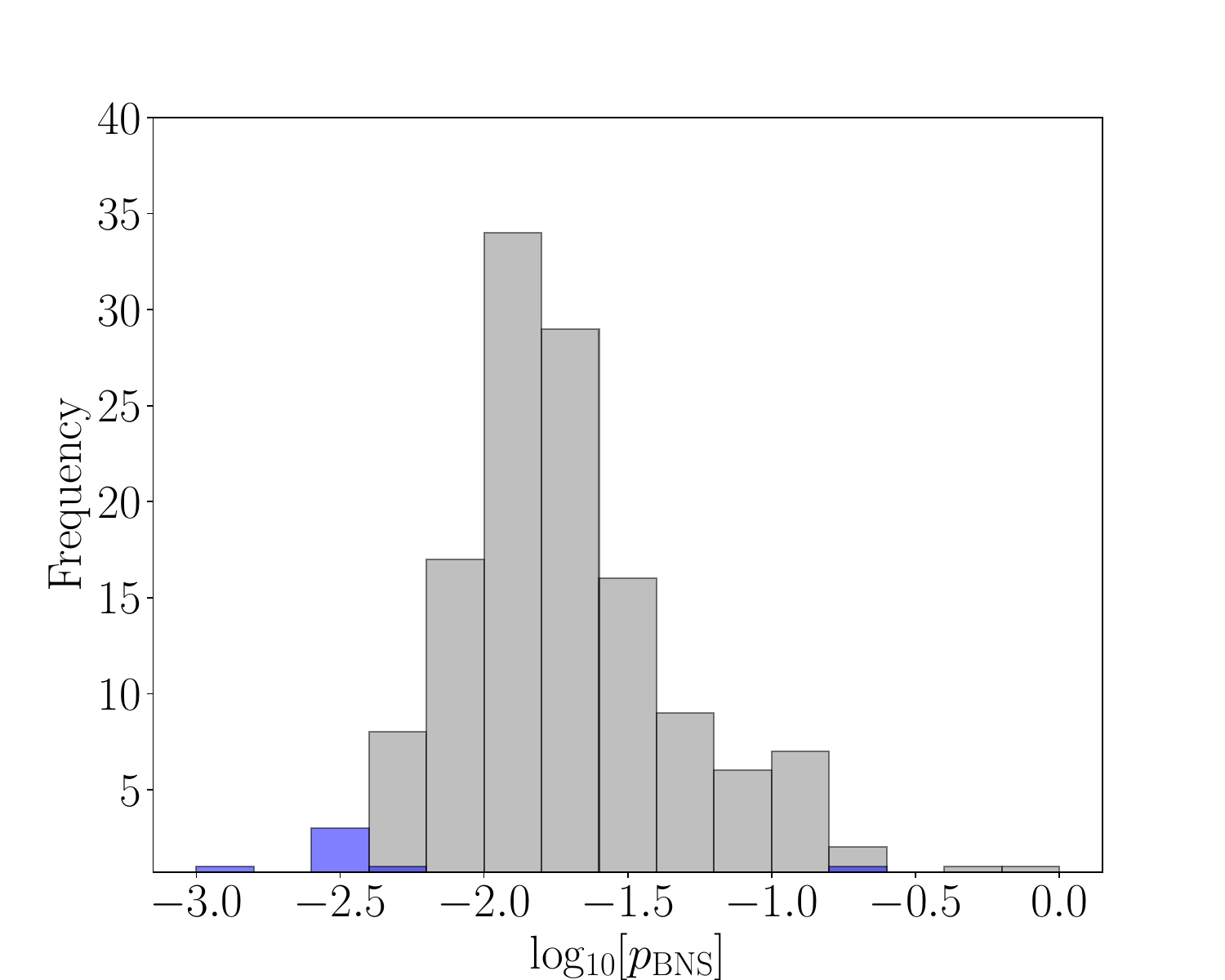}
    
    \caption{Distribution of the $p_{\rm astro}$ values for the CBC triggers detected during O3, with available GUANO data dumps. We distinguish with different colors the triggers received in low latency and the ones confirmed by offline analysis.}
    \label{fig:dist}
\end{figure*}

\begin{figure*}
    \centering
    \includegraphics[scale=0.6,trim={0.2cm 1.1cm 0.2cm 0.4cm},clip]{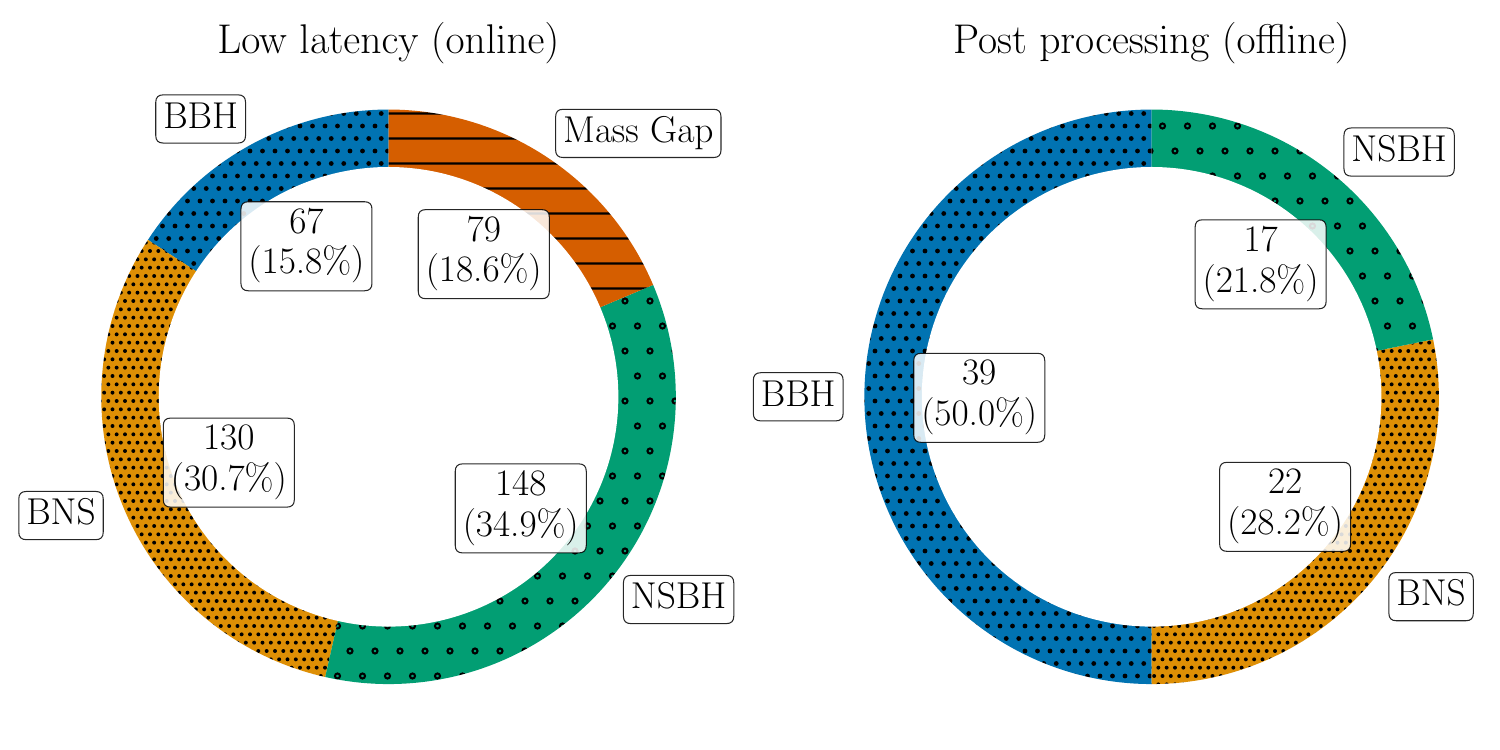}
    \caption{Left: distribution of the CBC triggers from O3 received in low latency, which had successful GUANO data dumps, divided in the BBH, NSBH, BNS, and Mass Gap classes. Right: analogous distribution for the CBC candidates confirmed by the offline analysis. In the post-processing, the Mass Gap classification has been subsumed into BBH and NSBH.}
    \label{fig:piechart}
\end{figure*}

In Figure~\ref{fig:dist} we show the histograms of the $p_{\rm astro}$ probabilities for all the low-latency CBC candidates processed by GUANO, for a total of 424 candidates, divided into 67 BBH, 130 BNS, 148 NSBH, and 79 Mass Gap events (Fig. \ref{fig:piechart}, left panel). In the offline post-processing of GW candidates, the Mass Gap classification was removed, classifying an object with $M>3M_{\odot}$ as a black hole. This implies that all the CBC events with at least one component mass in the range $[3-5]M_{\odot}$ are re-distributed into the BBH and NSBH classes. The distribution of the updated $p_{\rm astro}$ classification released by LVK is over-plotted in Figure~\ref{fig:dist}, and the classes are divided in 39 BBH, 22 BNS and 17 NSBH candidates (Fig. \ref{fig:piechart}, right panel).

\subsection{GW sky localization}
For the selection of the GW sky localization for each candidate we adopt the following scheme:
\begin{enumerate}
    \item Above--threshold GW candidate: The GW candidate is contained in the list of high-probability ($p_{\rm astro}>0.5$) candidates reported in Table~2 of GWTC-2.1 \citep{GWTC-2.1PhysRevD.109.022001} or Table~1 of GWTC-3 \citep{GWTC3-2023PhRvX}. The GW sky localizations are downloaded from the parameter estimation data releases of GWTC-2.1 and GWTC-3.\footnote{GWTC-2.1 release \url{https://doi.org/10.5281/zenodo.6513631}, GWTC-3 release \url{https://doi.org/10.5281/zenodo.8177023}} We use the results derived from a combination of IMRPhenomXPHM \citep{Pratten2021PhRvD} and SEOBNRv4PHM \citep{Ossokine2020PhRvD} waveforms (labeled as `Mixed' in the release). 
    \item Subthreshold GW candidate: The GW candidate is classified as low-probability ($p_{\rm astro}<0.5$) in the offline analysis. The GW sky localization is produced by BAYESTAR \citep{Sing2016,Singer2016ApJ} and is taken from the GWTC-3 release, which contains both O3b events and updated O3a events. If multiple events are present for a single GW candidate, the pipeline with the highest $p_{\rm astro}$ is considered for the selection of the sky localization.
    \item Non-confirmed low-latency GW candidate: The GW candidate has an associated low-latency alert, but the event has not been confirmed by the offline analysis. The GW sky localization is downloaded from GraceDB, selecting the preferred event.
\end{enumerate}

\section{Targeted searches using GUANO-NITRATES}\label{sec:target}

Targeted searches are carried out in real time for all types of transients such as GRBs, FRBs, neutrino events as well as GW triggers, on the event mode data obtained using GUANO. The targeted search pipeline that is currently operational is NITRATES. This is a maximum--likelihood framework that forward models signals through the entire instrument response \citep{NITRATES2022}. The BAT responses are created by simulating the photon paths through all detector segments using Geant4, which is a particle-interaction simulator software toolkit \citep{AllisonGEANT2016}. Unlike the standard BAT responses, the NITRATES responses account for all the detectors on the focal plane, regardless of their coding by the mask. The responses also encode details on the gamma-ray interactions inside the instrument, which carry additional information. This approach enables substantial sensitivity gain compared to the conventional technique of cross-correlation imaging, which translates to a 50\% increase in the detection horizon distance for a GRB 170817A-like burst compared to the onboard imaging. Details of the NITRATES response generation and the analysis pipeline are provided in \citet{NITRATES2022}.

A GRB-like transient signal is described using the sky localization and parameters that are specific to the assumed spectral model (the peak energy and spectral slope). This framework then computes the significance of each signal using a test statistic (TS) by maximizing the log-likelihood (LLH) as a function of signal parameters. The likelihood ratio test statistic $\Lambda$ is used to compare the source signal+background model (described using a set of parameters, $\Theta_{\text{sig}}$ that maximizes the LLH) to a background-only model (described by the set of parameters $\Theta^{\text{off}}_{\text{bkg}}$, that maximizes the LLH in the off-time window) and is defined as follows \citep{NITRATES2022}:
\begin{equation}
    \Lambda = -2[\rm LLH(\Theta^{\text{off}}_{\text{bkg}} | N^{\text{on}}) - LLH(\Theta_{\text{sig}}, \Theta^{\text{off}}_{\text{bkg}}| N^{\text{on}})],
\end{equation}
where $N^{\text{on}}$ is on-time data.

The search pipeline workflow can be summarized as follows: 
\begin{enumerate}
    \item The event mode data is cleaned and filtered to discard potential glitches and artifacts from cosmic rays, and to flag poorly behaving detectors. Good time intervals (GTIs) are determined, where there is quality data for the analysis.
    \item A time window of 50~s (from the pre- and post-trigger intervals) is identified as the background interval. It is then utilized to model contributions to the background from known bright sources and diffuse sources.
    \item To narrow down the search parameter space, a set of simple analyses are performed to select a list of interesting start times and durations (hereafter referred to as time seeds) as well as portions of the BAT FOV (position seeds). 
    \item Finally, the log-likelihoods are computed for all parameters corresponding to the shortlisted time and position seeds.
\end{enumerate}
In essence, the set of signal parameters that maximizes the log-likelihood is the most preferred set of parameters. 

The NITRATES likelihood analysis outperforms the onboard mask-weighted imaging analysis by delivering superior sensitivity, given the increased effective area (see Fig. 2 in \citealt{NITRATES2022}). At the cost of a significantly increased computational time, this method is capable of delivering arcminute scale localization for events that fall inside the BAT FOV, even when the transient event does not trigger \textit{Swift}-BAT onboard (\citealt{GRB211106AGCN2021,GRB221231AGCN2023A, GRB220107AGCN2022}). The NITRATES pipeline has the ability to distinguish between bursts that come from in and outside the BAT FOV. NITRATES has also accurately localized sufficiently bright bursts outside the FOV \citep{NITRATES2022}.

\textit{Swift}-BAT GUANO was operating during the O3 and was successfully procuring event mode information in response to GW subthreshold triggers \citep{GUANOTohuvavohu2020}. We describe the targeted search analysis that has been carried out using the NITRATES version 0.0.1 which was available in early 2022.\footnote{https://github.com/Swift-BAT/NITRATES/tree/py2} The targeted search analysis that was operational in O3 corresponded to a preliminary version of the NITRATES code, that has since undergone several stages of development. The most updated version is publicly available on GitHub.\footnote{https://github.com/Swift-BAT/NITRATES}  

During O3, for a total of 636 GW triggers, GUANO dumped either 200~s or 90~s of event mode data, for public triggers and for privately communicated triggers, respectively. The choice of the width of the temporal window is made to avoid an overload of downlink data in the process of GUANO data dump. The targeted search pipeline was run in a time window of $\pm$20~s centered around the trigger time. The search was carried out on 8 time bins (0.128~s, 0.256~s, 0.512~s, 1.024~s, 2.048~s, 4.096~s, 8.192~s and 16.384~s) and 9 energy bins (between 15--350~keV). The results from the search are reported using the following set of parameters: 1) the maximum $\sqrt{\text{TS}}$  describes the statistical significance of a potential detection (see Section \ref{nit_res}); 2) $\Delta \rm LLH_{out}$ indicates the preference of the search to a location inside or outside the BAT FOV, and 3) $\Delta \rm LLH_{peak}$ indicates the confidence of the search in localizing the source to arcminute scales. The maximum $\sqrt{\text{TS}}$ is empirically mapped to a FAR using the distribution of $\sqrt{\text{TS}}$ values found from analyzing 51 ks of random data (see sec. 7.6 in \citealt{NITRATES2022}). 

The NITRATES search was performed on the ROAR supercomputing cluster on a set of 200 virtual cores for a total of $\sim 600 \times 800$ CPU hours for the entire GW sample.

\section{Results from NITRATES}
\label{nit_res}
The targeted search analysis provides a list of top candidates whose spatial, temporal, and spectral parameters maximize the log-likelihood. In order for a candidate to be qualified as a confident detection, we require that the resulting detection significance parameter $\sqrt{\text{TS}}$ must exceed the threshold value of 8, corresponding to a FAR $\sim 4 \times 10^{-5}$ Hz. Being a targeted search, the NITRATES analysis can give a false positive with $\sqrt{\text{TS}}>8$ with a probability which follows a Poissionian distribution:
\begin{equation}
    P(N_{\text{det}}\geq 1)=1-P(N_{\text{det}}=0)= 1-e^{-\text{FAR} \times \Delta t},
\end{equation}
with $\Delta t=40$ s being the width of the search window. This leads to a pre-trial p-value of $1.6\times 10^{-3}$. Since the NITRATES analysis is performed on all GW triggers with a FAR $<2$~day$^{-1}$, and considering that there are $N_{\rm GW-search}=5$ independent GW pipelines, the rate of expected false positive candidates falling within the temporal search window around a GW trigger with $\sqrt{\text{TS}}>8$ is $\sim 5 \times 2 \times 1.6\times 10^{-3} $ /day $\sim $ 1/(60 day).\footnote{Since the GW pipelines are not totally independent, a realistic value of $N_{\rm GW-search}$ is likely below 5, leading to an overall rate of false NITRATES candidate below 1/(60~day).}

For the entire sample of 636 low-latency triggers processed using NITRATES, we have no candidates that qualify as detection of a signal of astrophysical origin. None of the top candidates within the $\pm 20$ s search window are coincident with the GW triggers. A temporal coincidence with a GW trigger is claimed if the NITRATES search finds a candidate with $\sqrt{\text{TS}}>8$ and $|t_0-t_{\rm start}|<20$ s, where $t_0$ is the GW trigger time and $t_{\rm start}$ is the starting time of the temporal bin with highest ranking statistics. A detailed list of all the NITRATES results for the entire sample analyzed during O3 is provided in Table \ref{low-latency}. We discuss specific false positive candidates in Section \ref{sec:falsepos}. 

If the GW trigger time is included in the time window corresponding to slew mode of BAT, the analysis cannot be performed using NITRATES since the targeted search requires stable attitude information to compute the background. Similarly, some triggers have insufficient exposure time, preventing the NITRATES analysis. In this case neither TS results nor flux upper limits can be computed. 
As a cut to narrow down the parameter space, the targeted search selects time seeds as described in Section \ref{sec:target}. If there are no time seeds that pass the preliminary cuts then there will be no final likelihood computations. Results for these types of triggers are indicated as NFL (No Final Likelihood) in Table \ref{low-latency}. In the case of NFL triggers, though, the flux upper limit can be computed, since a full likelihood analysis is not required.

\subsection{False positives}\label{sec:falsepos}

We did not find any candidate associations from any of the triggers with BAT. However, the targeted search pipeline did result in the detection of six candidates with a significance above the NITRATES detection threshold of $\sqrt{\text{TS}}=8$. These candidates were examined to understand our false positive population. 
S200327j ($\sqrt{\text{TS}}\sim 22$), S200324ax ($\sqrt{\text{TS}}\sim 11$), and S200225af  ($\sqrt{\text{TS}}\sim 10.5$) are triggers that occurred during the passage of \textit{Swift} in the proximity of the South Atlantic Anomaly (SAA). The background characterization becomes unreliable when the spacecraft is either entering or leaving the SAA on account of increased background contamination. Analyses close to the SAA are not considered to be during good data times and are not accounted for in the NITRATES FAR distribution. The pipeline expanded into these not good data times to search for any exceptional signals. 

S200130ai corresponds to a subthreshold GW trigger at $T_0=$ 2020-01-30T09:59:58 that was identified by the CBC search as a NSBH candidate with a $p_{\rm astro}= 0.008$ and a GW FAR $\sim 1.8 \times$10$^{-5}$~Hz. It was detected using NITRATES at a significance of $\sqrt{\text{TS}}\sim16.3$  with a $\Delta \rm LLH_{out}=-19.68$ and $\Delta \rm LLH_{peak}=2.14$, consistent with a sky localization outside the BAT-FOV. The highest log-likelihood candidate, was identified to arise 1.5 s prior to the GW trigger time. Due to the low value of $\rm \Delta LLH_{peak}$, we do not have an arcminute-level precision on the sky localization. The candidate was associated with a \textit{Fermi} trigger 602071201 (GCN 26944, \citealt{FermiGRB2020GCN}) and was classified as a long GRB. The Fermi localization is RA $= 137.5 $ deg, Dec $= -51.3$ deg, with a statistical uncertainty of 3.5 degrees.  The Interplanetary Gamma-Ray Burst Timing Network (IPN) further localized the event in a 3-sigma error box with an area of 1487 arcmin$^2$ and centered at an RA $=134.742$ deg and Dec $=-49.627$ deg \citep{2020GCN26949IPN}. Although this event presents a temporal coincidence with the GW trigger, on account of the lack of spatial coincidence with the GW location, this event is discarded from being associated with the GW subthreshold trigger. Additionally, this low-latency GW candidate has not been confirmed by offline analyses.

In S190919au, a peculiar dip ($\sim$20~s) in the background may have contributed to a false elevation in the signal detection statistic, by causing an under representation of the background rate. For S190919u, we obtain a $\sqrt{\text{TS}}\sim 8.0$, which corresponds to a FAR of $\sim 4 \times 10^{-5}$ Hz. Including the two events confirmed to be unrelated to their corresponding GW trigger, the search identified three events with $\sqrt{\text{TS}}>8$ during good data times. This is compatible with the Poissonian error bars of the expected number of false positives, which corresponds to $(4 \times 10^{-5}$~Hz)~$\times (40$~s$)\times$ 636 GW triggers $\sim 1$. Additionally, there is substantial uncertainty on the empirically derived NITRATES FAR at $\sqrt{\text{TS}}\gtrsim 8.0$ due to low statistics (see Fig. 16 in \citealt{NITRATES2022}). Including the finite statistics used to determine the NITRATES FAR, the 90\% confidence interval on the expected number of false positives is 0.1 - 3.1.

\subsection{Computation of flux upper limits}

\begin{figure*}
\centering

\includegraphics[trim=110 150 120 150, clip, width=0.88\columnwidth]{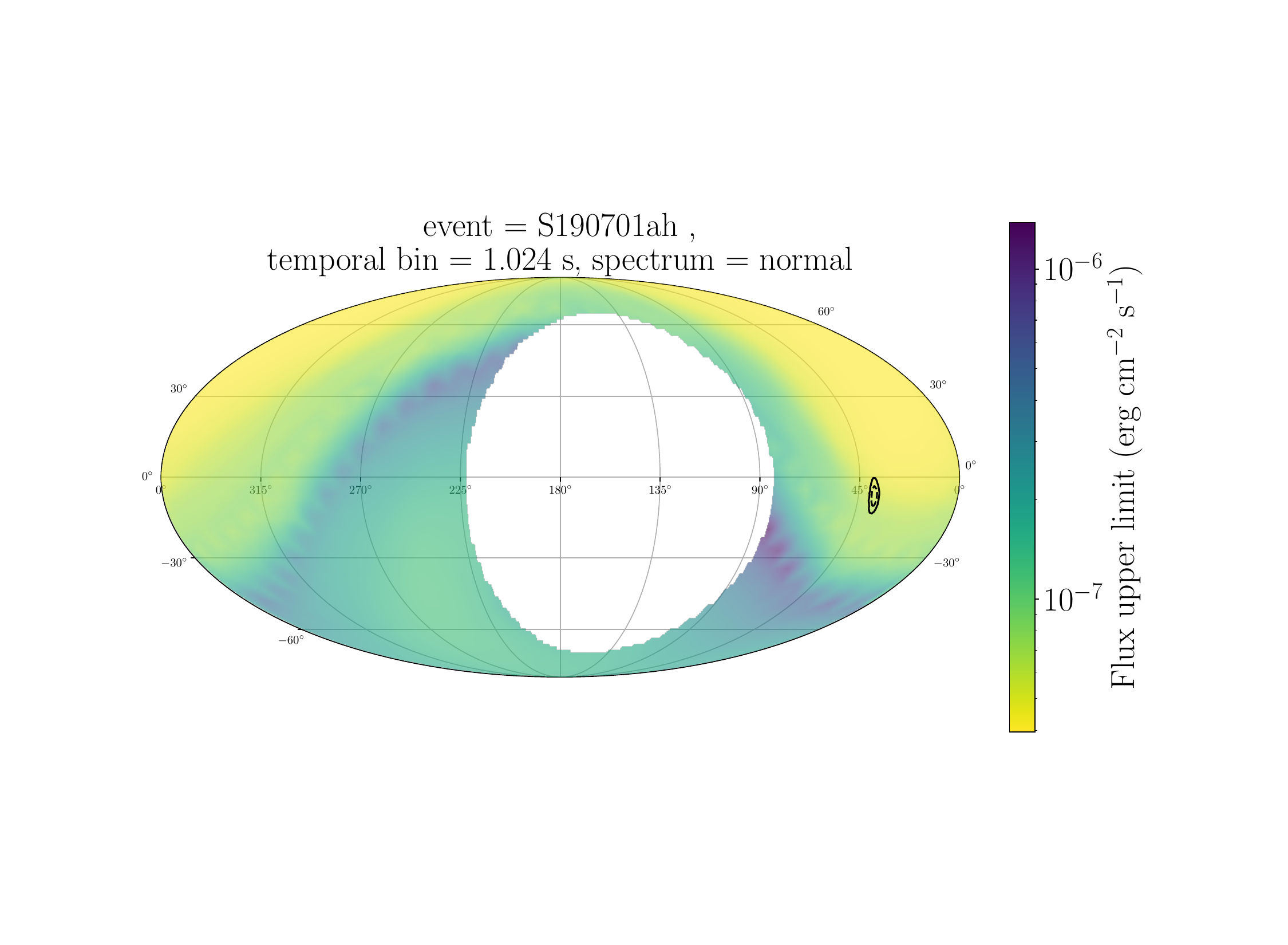}
\hspace{3em}
\includegraphics[trim=110 150 120 150, clip, width=0.88\columnwidth]{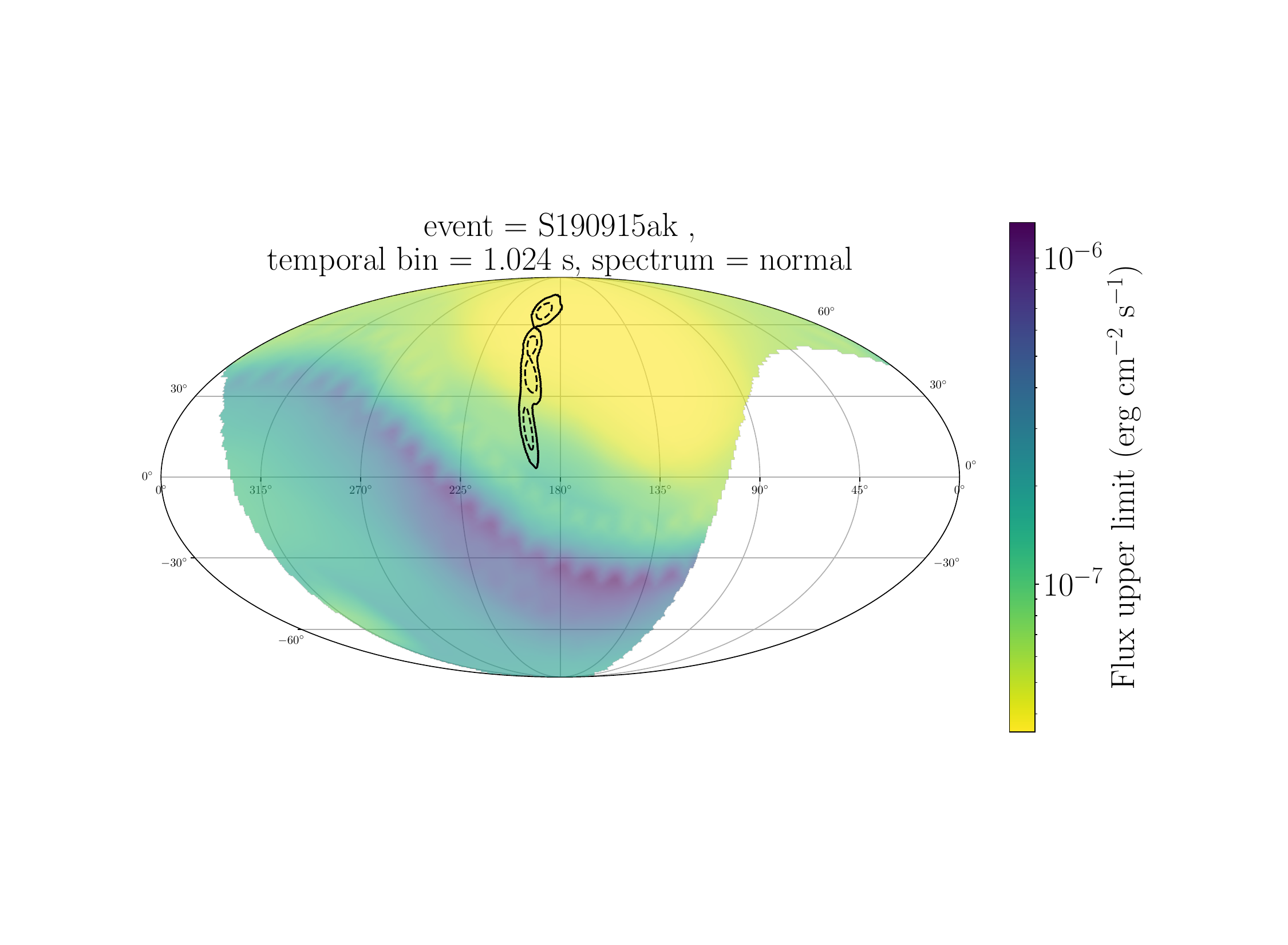}
\vspace{0.5cm}
\includegraphics[trim=110 150 120 150, clip,width=0.88\columnwidth]{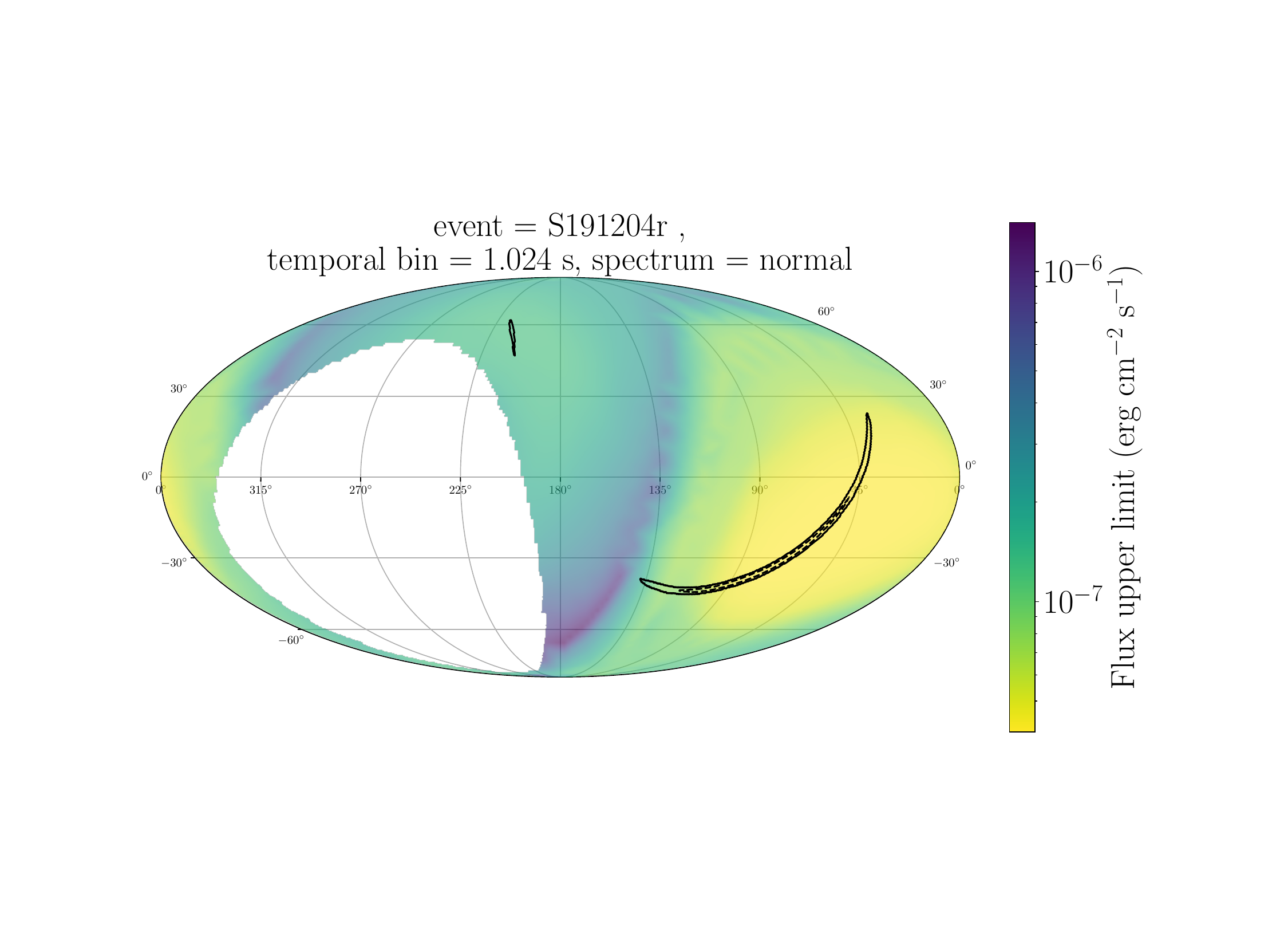}
\hspace{3em}
\includegraphics[trim=110 150 120 150, clip,width=0.88\columnwidth]{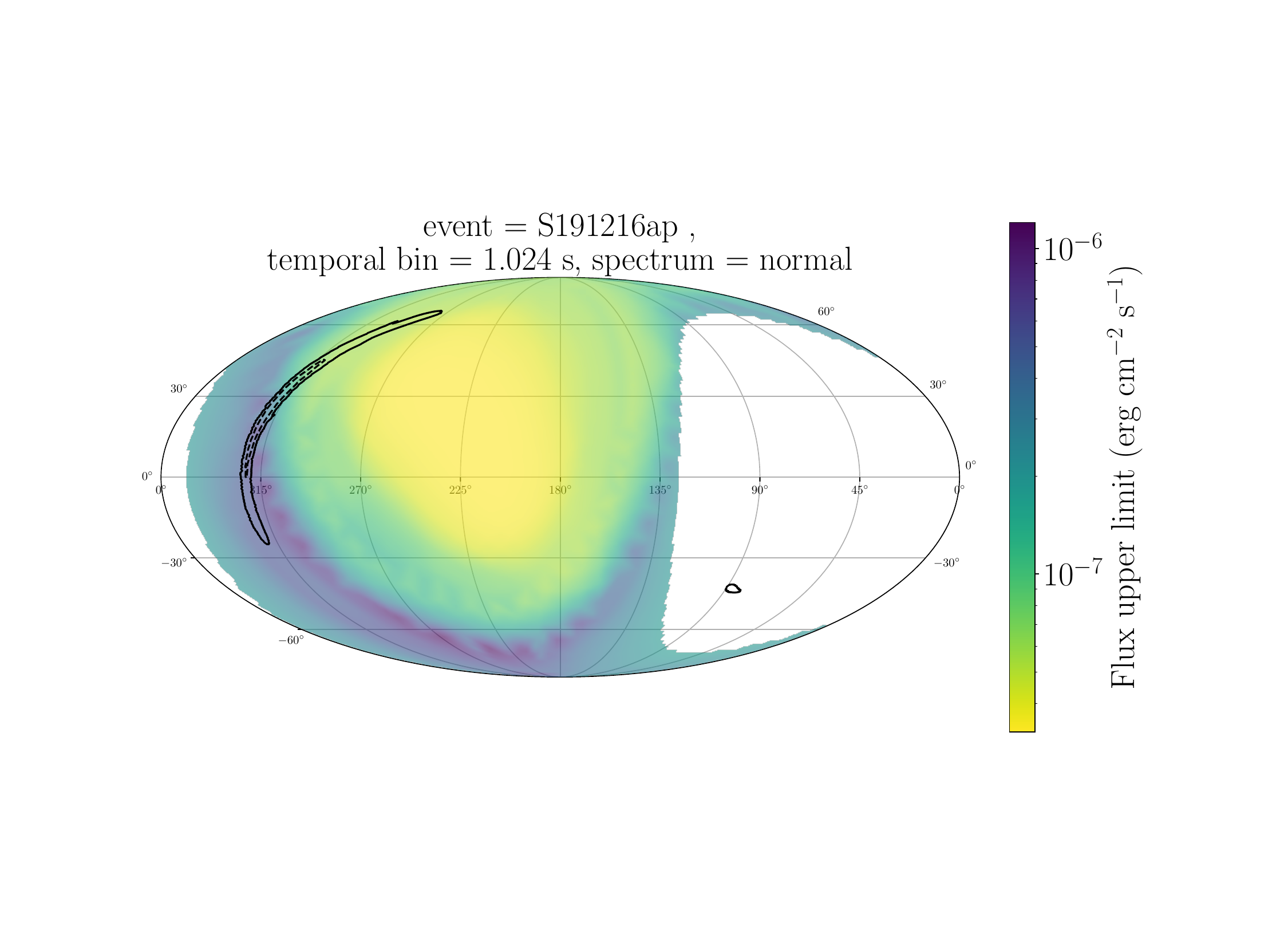}
\vspace{0.5cm}
\includegraphics[trim=110 150 120 150, clip,width=0.88\columnwidth]{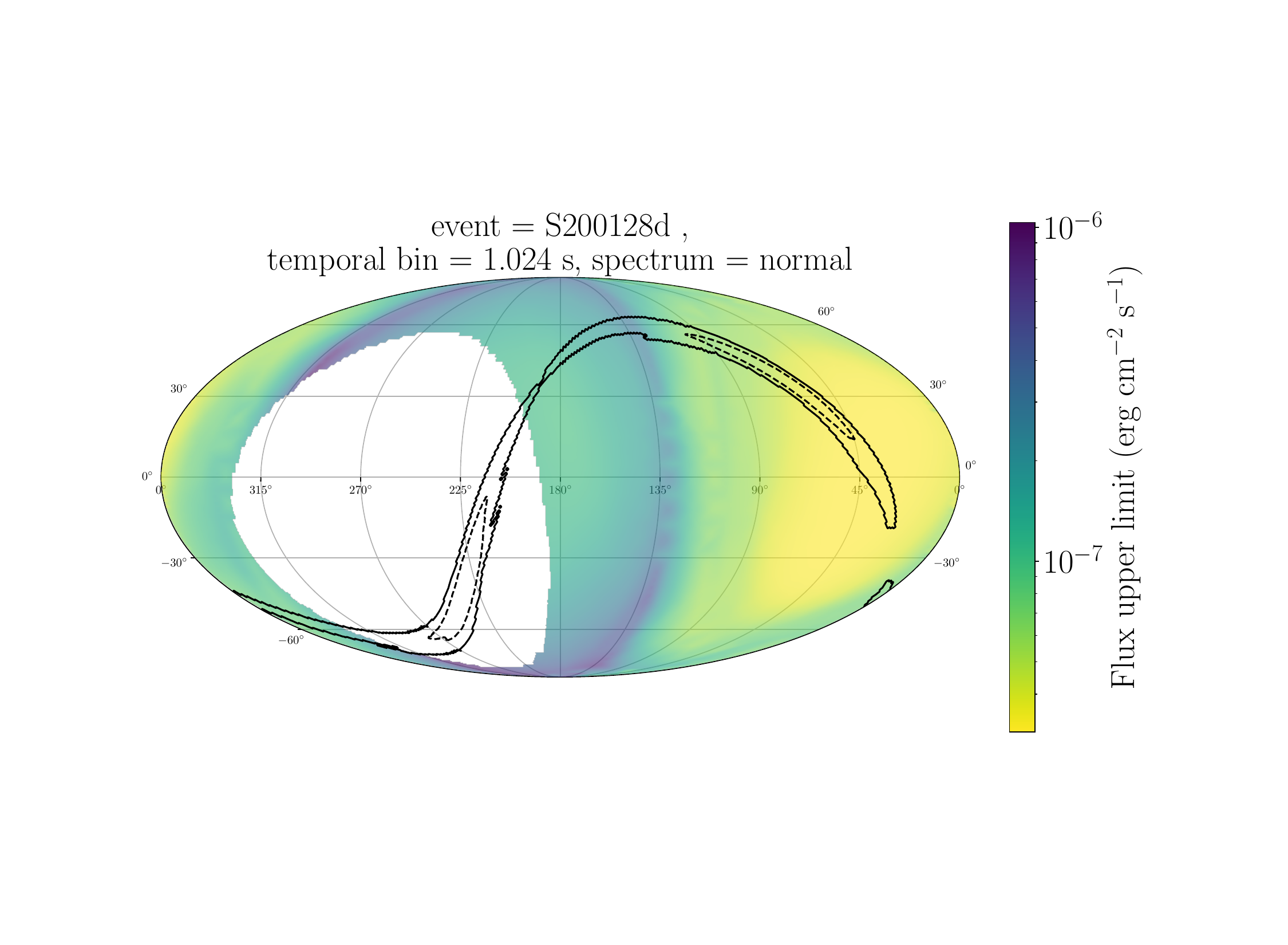}
\hspace{3em}
\includegraphics[trim=110 150 120 150, clip,width=0.88\columnwidth]{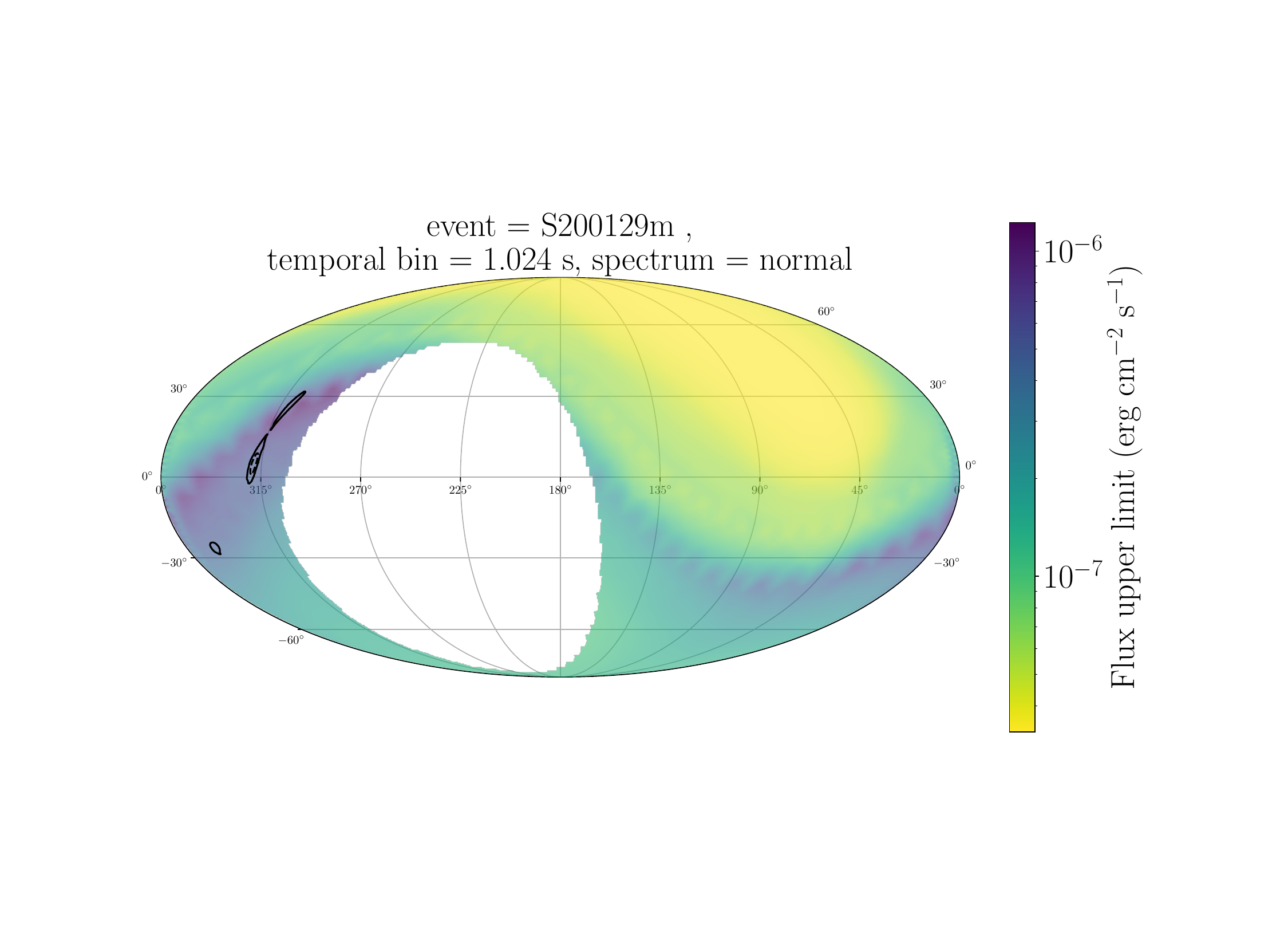}

\caption{Flux upper limit maps are shown for all the O3 catalog events with a $p_{\rm astro}>0.5$ that were processed successfully using NITRATES. The color bar indicates the upper limit in the 15--350~keV \textit{Swift}-BAT band as a function of the sky position. The part of the sky in white corresponds to the area covered by the Earth. The solid and dashed contours are the GW 90\% and 50\% credible levels, respectively.}

\label{fig:ul1}
\end{figure*}

\begin{figure*}
\centering
\includegraphics[trim=110 150 120 150, clip,width=0.88\columnwidth]{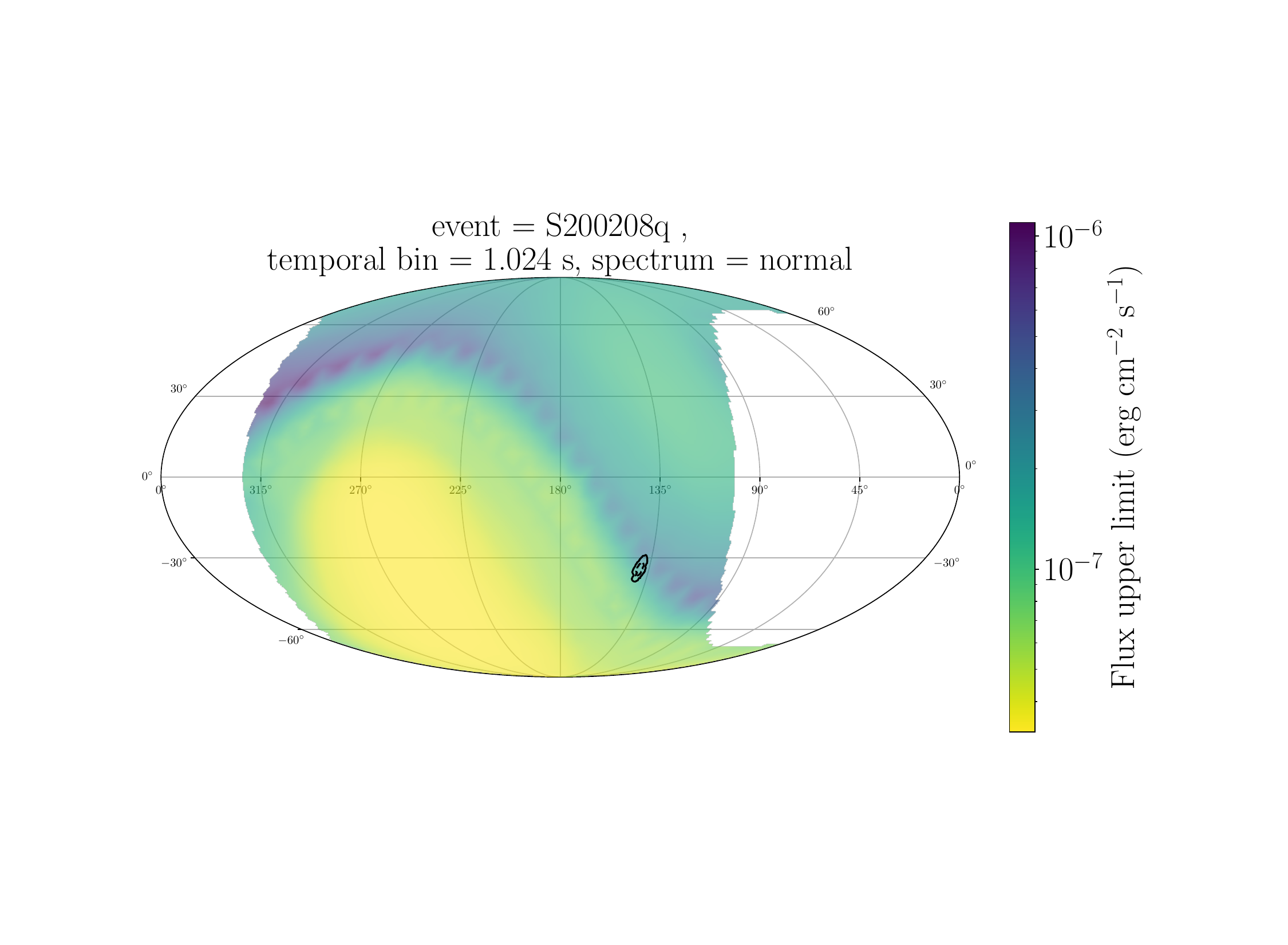 }
\hspace{3em}
\includegraphics[trim=110 150 120 150, clip,width=0.88\columnwidth]{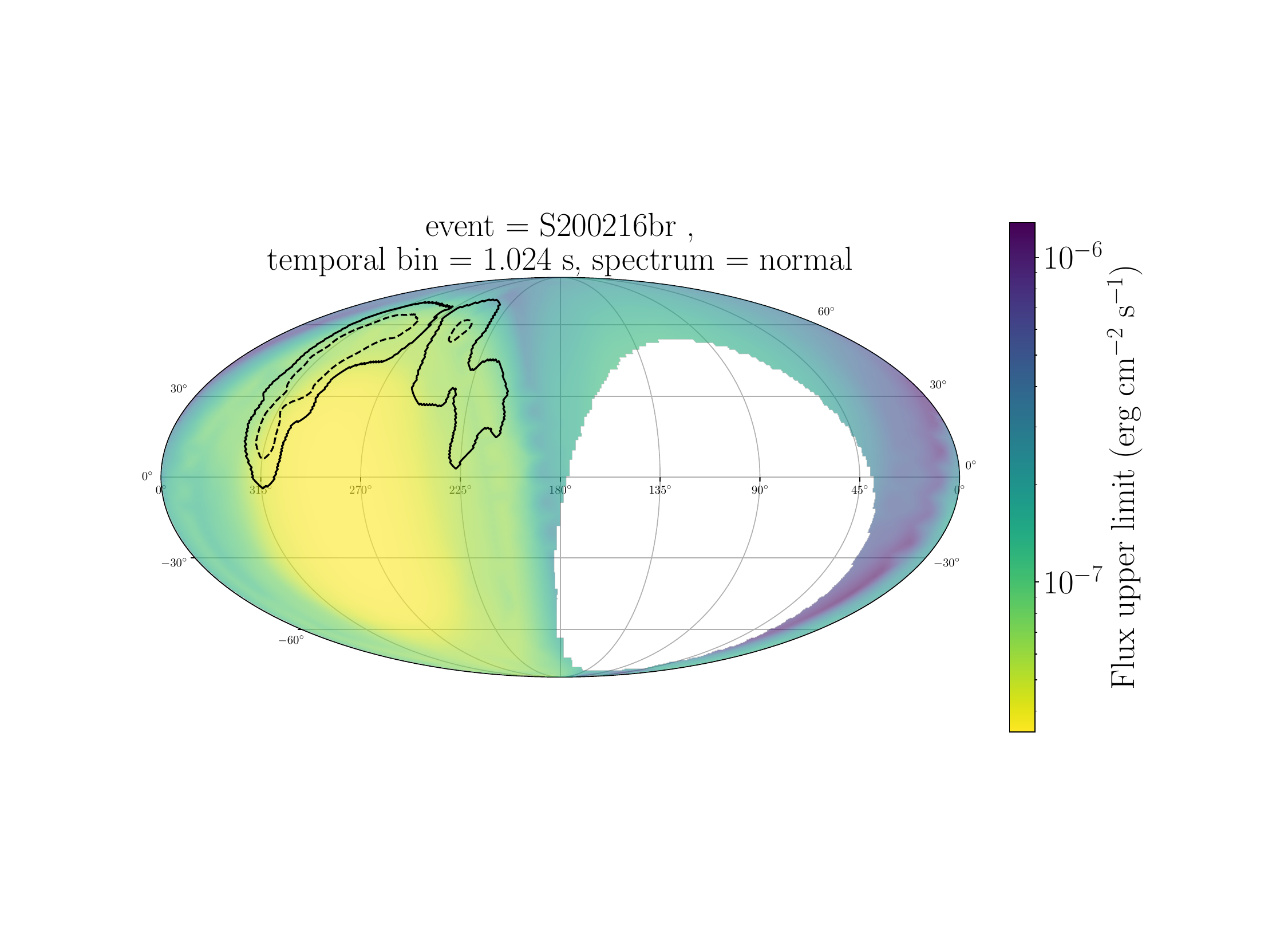} 
\vspace{0.5cm}
\includegraphics[trim=110 150 120 150, clip,width=0.88\columnwidth]{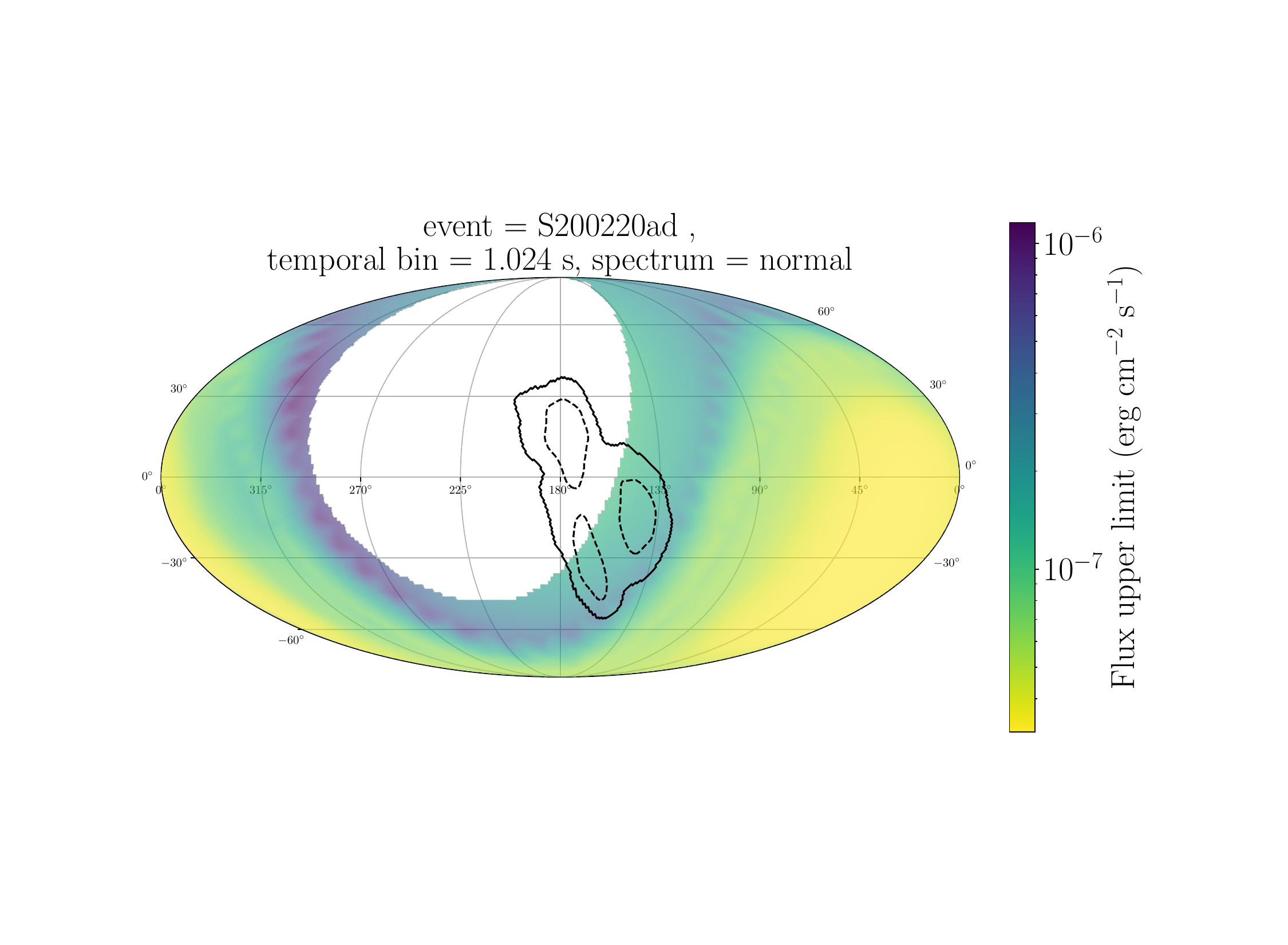}
\hspace{3em}
\includegraphics[trim=110 150 120 150, clip,width=0.88\columnwidth]{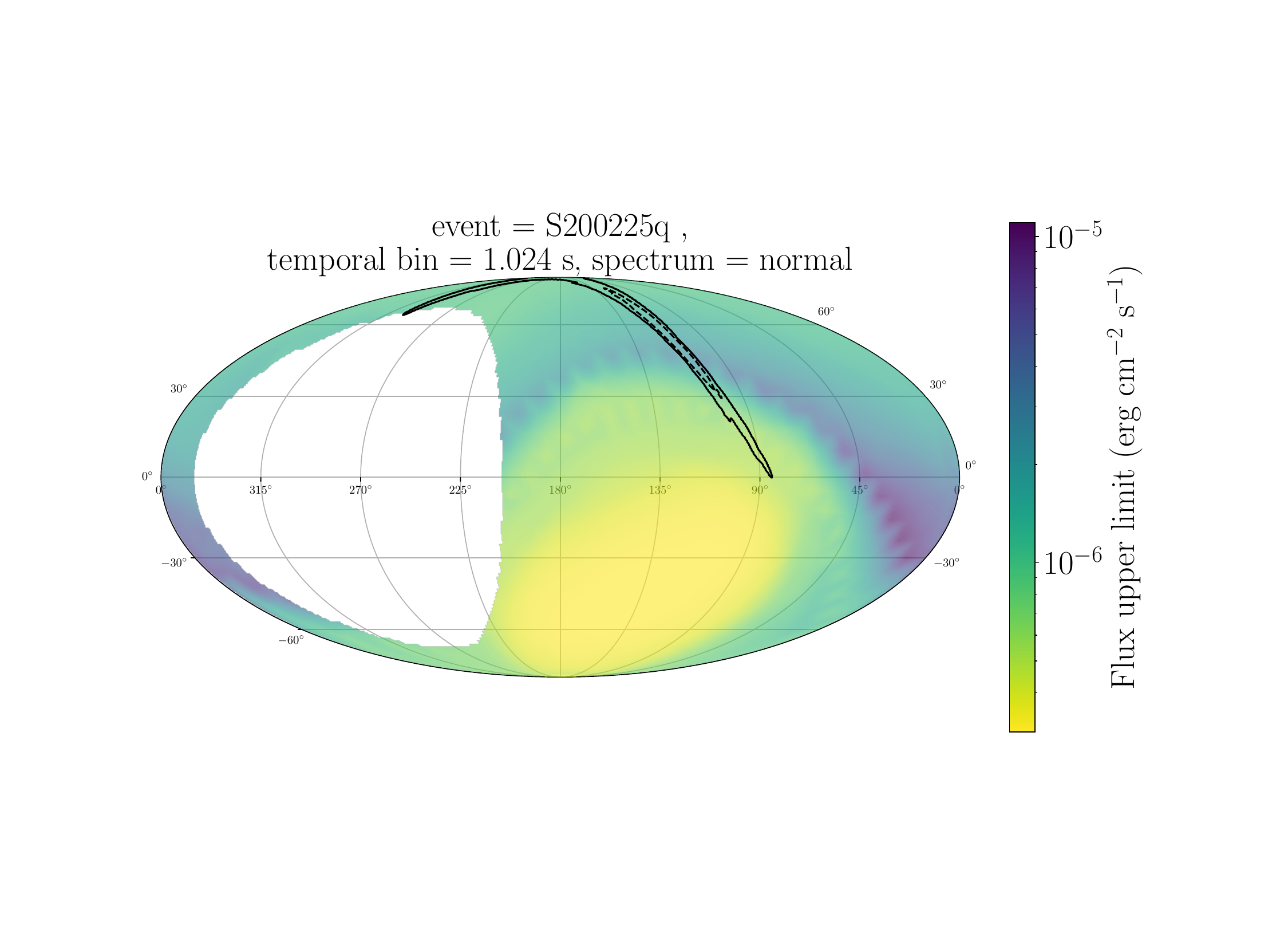}
\vspace{0.5cm}
\includegraphics[trim=110 150 120 150, clip,width=0.88\columnwidth]{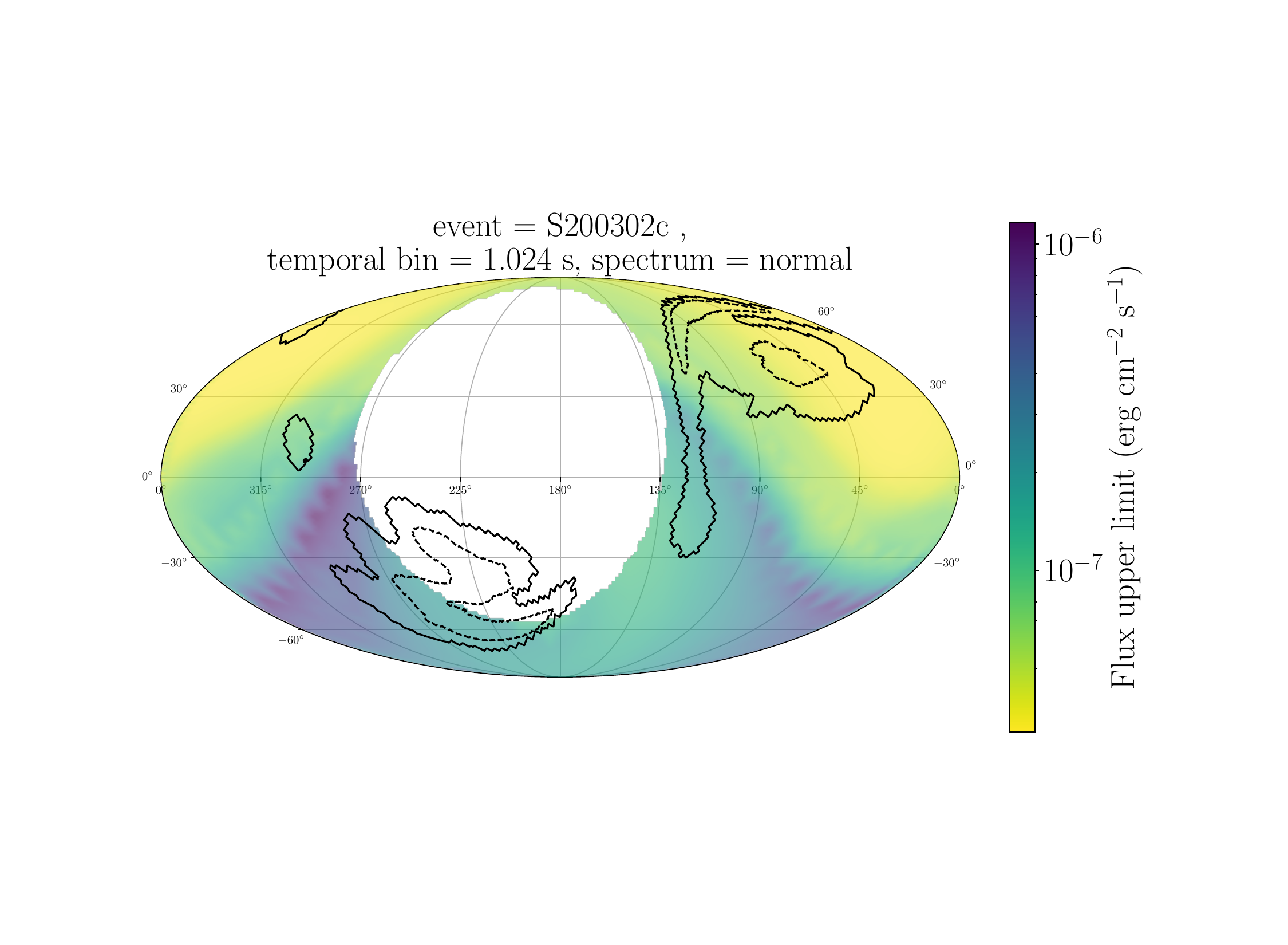}
\hspace{3em}
\includegraphics[trim=110 150 120 150, clip,width=0.88\columnwidth]{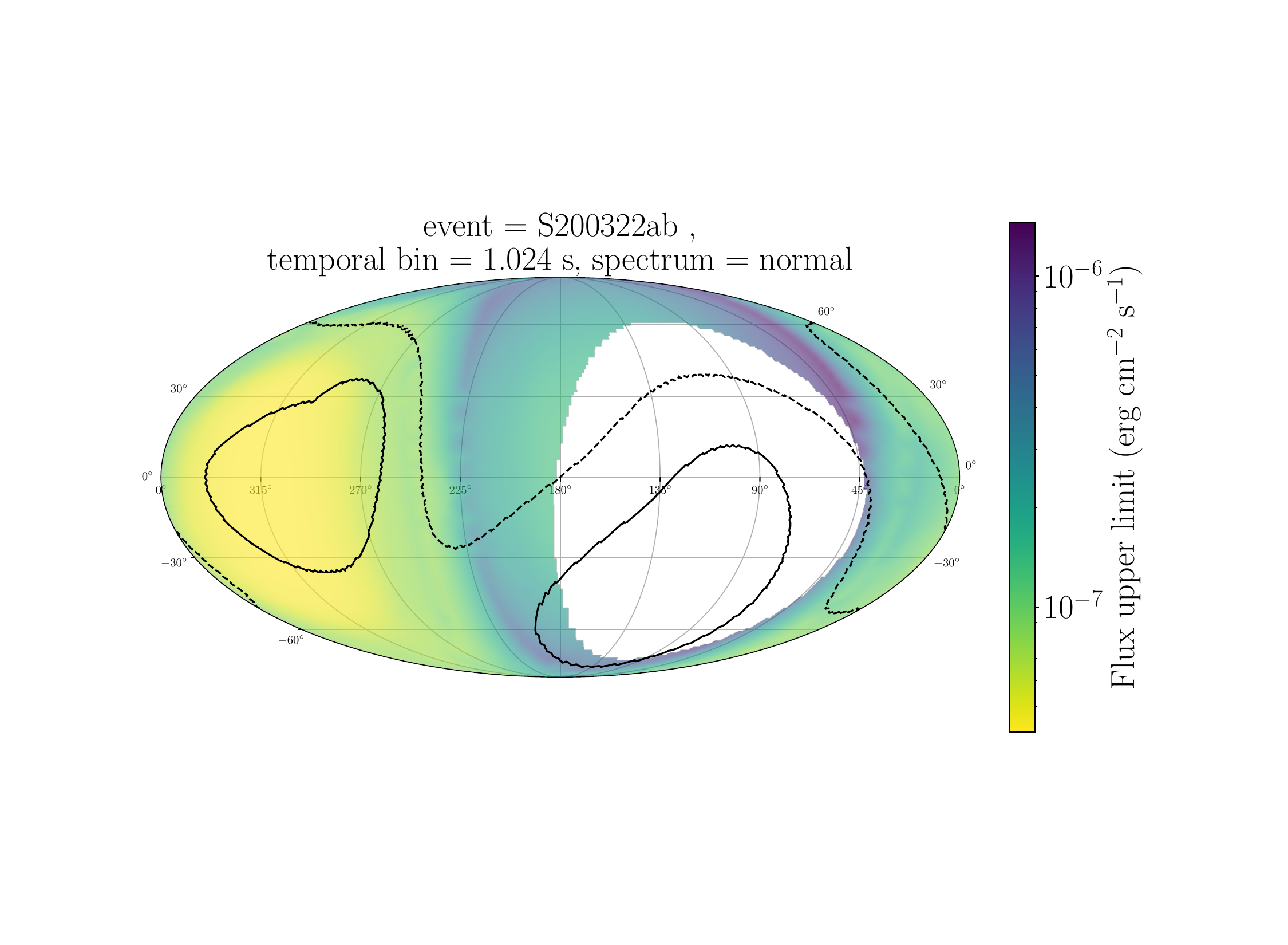}



\makebox[0.8\textwidth]{\centering \textbf{Figure 3.} (continued)}

\label{fig:ul2}
\end{figure*}

\begin{figure*}
    \centering
    \includegraphics[width=1.7\columnwidth]{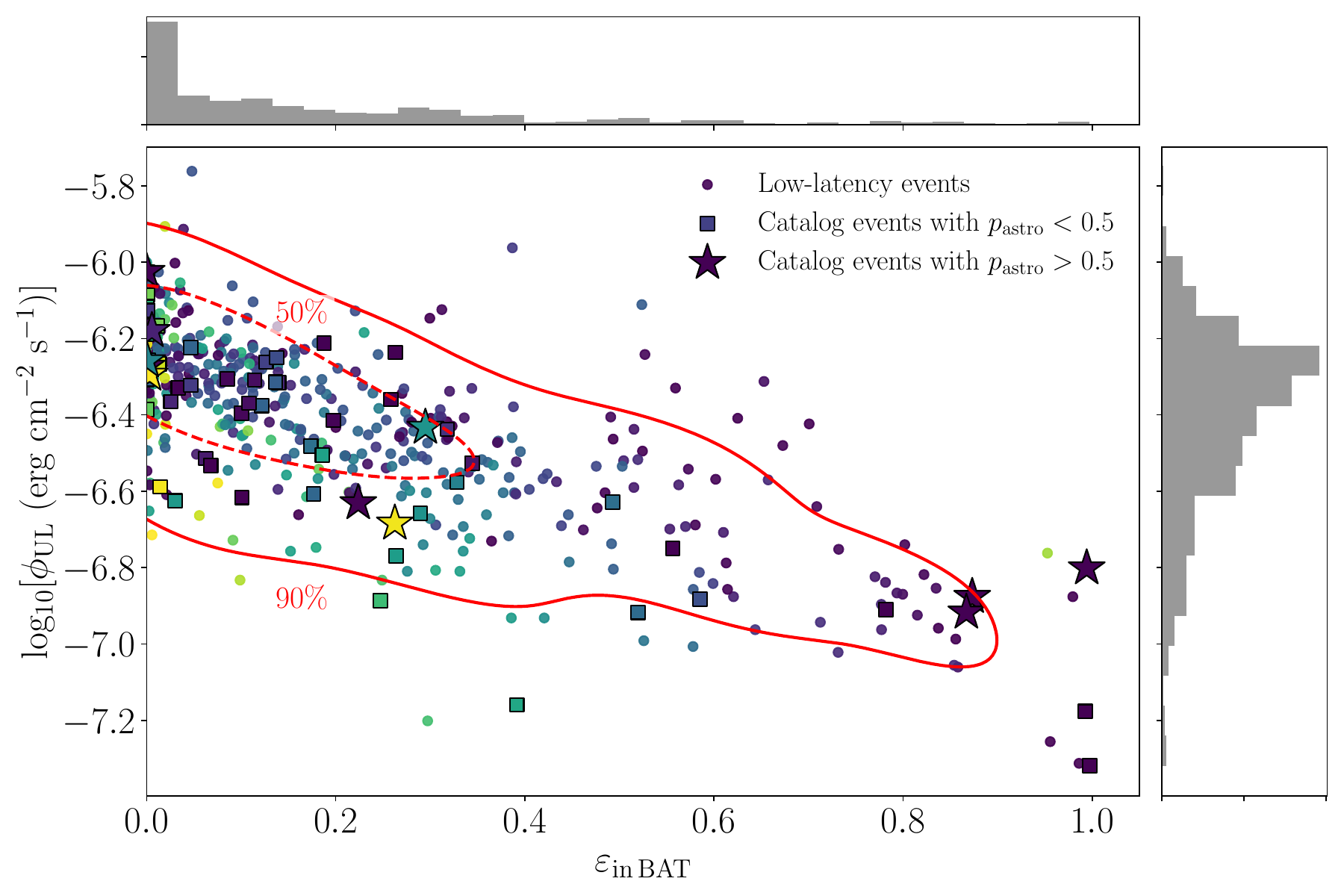}
\hspace{1em}
   \includegraphics[width=0.22\columnwidth]{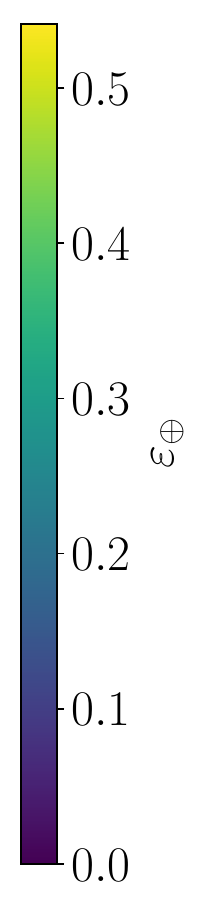}
   
    \caption{The 15--350~keV flux upper limit $\phi_{\rm UL}$ derived with NITRATES as a function of $\varepsilon_{\rm in \, BAT}$, namely the probability that the GW candidate is contained inside the BAT FOV. The plot includes all the GW candidates received during O3, including both Burst and CBC events. With different symbols we distinguish the GW candidates received in low latency and the ones confirmed and included in the O3 catalog, separated in $p_{\rm astro}>0.5$ and $p_{\rm astro}<0.5$. The dashed and solid red lines are the 50$\%$ and 90$\%$ containment regions of the scatter plot, respectively. The one-dimensional histograms of $\phi_{\rm UL}$ and $\varepsilon_{\rm in \, BAT}$ are reported on the sides. The color bar indicates the value of $\varepsilon_{\Earth}$, the probability that the GW candidate is occulted by the Earth.}
    \label{fig:gw_in}
\end{figure*}             

Since each GW trigger processed in this analysis resulted in a non-detection in \textit{Swift}-BAT, we estimate the flux upper limits in the following manner. The NITRATES analysis generates rates curves in the 15--350~keV energy band from the GTIs of the filtered event list. The number of active detectors corresponding to each trigger is read out from its respective detector mask file. A linear fit is carried out to the background window of duration 50~s. We then estimate the 5$\sigma$ count rate and the corresponding uncertainty over the full signal window which has a $\pm$20~s duration. This is computed for all the 8 time bins (see Section \ref{sec:target}). We further convert the 5$\sigma$ count rates to flux upper limits, as a function of sky position, in the 15--350~keV band, by convolving different spectral models with the NITRATES responses for each time bin iteration. We select 929 grid points on the sky and interpolate upper limit values for locations in between. We assume the following different spectral templates:
\begin{enumerate}
    \item Band function \citep{band1993ApJ...413..281B} with a soft template ($E_{\rm peak}=70 $ keV, $\alpha=-1.9$, $\beta=-3.7$)
    \item Band function with a normal template ($E_{\rm peak}$ = 230 keV, $\alpha=-$1.0, $\beta=-$2.3)
    \item Cutoff power law function with a hard template ($E_{\rm peak}=$ 1500 keV, $\alpha=$ 1.5)
    \item Cutoff power law function that has been used to describe GRB 170817A ($E_{\rm peak}$ = 185 keV, $\alpha$ = 0.62) \citep{Goldstein2017170817}
\end{enumerate}
The parameters $\alpha$ and $\beta$ correspond to the low-energy and high-energy photon indices of the spectrum, respectively. The first three spectral templates are identical to the ones that are routinely adopted by \textit{Fermi}-GBM \citep{Goldstein2016-template}. In the rest of the paper, all the results are reported assuming a Normal spectral template.

Calling $\Omega=(\rm RA, Dec)$ the coordinates variable, for each temporal bin and spectral template we convert the upper limit map $\phi_{\rm UL}(\Omega)$ into a unique marginalized upper limit value:
\begin{equation}
\label{phi_ul}
    \phi_{\rm UL} = \int_{\Omega \notin \Omega_{\Earth}} \phi_{\rm UL}( \Omega) P_{\rm GW}(\Omega) \rm d \Omega,
\end{equation}
where $P_{\rm GW}(\Omega)$ is the posterior probability distribution of the GW sky position. The notation $\Omega \notin \Omega_{\Earth}$ means that the integral is limited to the region of the sky not occulted by the Earth. 
We report in Table \ref{low-latency} the marginalized flux upper limits for a 1 s time bin, assuming the normal spectral template. 
In Fig. \ref{fig:ul1}, we provide the sky maps reporting both the flux upper limits as a function of sky position and the GW contours (50$\%$ and 90$\%$ credible levels) for the GW candidates with $p_{\rm astro}>0.5$. 

As additional information, we also report the quantity $\varepsilon_{\rm in \, BAT}$, which quantifies the probability that the GW source is inside the BAT coded FOV and corresponds to
\begin{equation}
    \varepsilon_{\rm in \, BAT} = \int_{\Omega   \in\Omega_{\rm in}} P_{\rm GW}(\Omega) \rm d \Omega,
\end{equation}
where the integral is limited to the solid angle $\Omega_{\rm in}$, namely the portion of the sky where the BAT partial coding fraction is larger than 0.01. The location of $\Omega_{\rm in}$, i.e.,  the BAT FOV, is identified by the yellow region in the sky maps of Figure \ref{fig:ul1}. The higher the $\varepsilon_{\rm in \, BAT}$, the better the BAT covering of the GW error region, and the more constraining the derived upper limit. The flux upper limits as a function of $\varepsilon_{\rm in \, BAT}$ for all GW trigger candidates is shown in Fig. \ref{fig:gw_in}. We also indicate with different markers the sample of low-latency triggers, the confirmed list of subthreshold candidates ($p_{\rm astro}<0.5$), and the above-threshold candidates ($p_{\rm astro}>0.5$). In Table \ref{low-latency} we also report the probability that the GW source is occulted by the Earth, defined as:
\begin{equation}
    \varepsilon_{\Earth} = \int_{\Omega   \in \Omega_{\Earth}}  P_{\rm GW}(\rm \Omega) d \Omega,
\end{equation}
where $\Omega_{\Earth}$ is the solid angle subtended by the Earth from the \textit{Swift} reference system.

\subsection{Computation of luminosity upper limits}
We further convert the flux upper limits into luminosity upper limits for all the GW triggers with available information about the distance posterior distribution, namely only triggers identified by CBC searches. The luminosity upper limit in the rest frame band 1~keV--10~MeV is estimated as
\begin{equation}
\label{l_ul}
    L_{\rm UL}=\langle4 \pi  D_{\rm L}^2  k \phi_{\rm UL}\rangle,
\end{equation}
where $D_{\rm L}$ is extracted from the posterior probability $P(D_{\rm L})$ reported in the GW sky localization files, while $k$ is the k-correction and corresponds to 
\begin{equation}
  k=  \frac{I[1 \,\mathrm{ keV}/(1+z),10 \, \mathrm{ MeV}/(1+z)]}{I[15 \,\mathrm{ keV}, 350 \,\mathrm{ keV}]},
\end{equation}
where 
\begin{equation}
  I[a,b]=\int_{a}^{b} E \frac{\mathrm{d} N}{\mathrm{~d} E}(E) \mathrm{d} E,  
  \end{equation}
and d$N$/d$E$ is the assumed photon spectrum. The band 1~keV--10~MeV is chosen since it is usually adopted to report the bolometric luminosity of GRBs.

\begin{figure*}
    \centering
    \includegraphics[width=0.5\columnwidth]{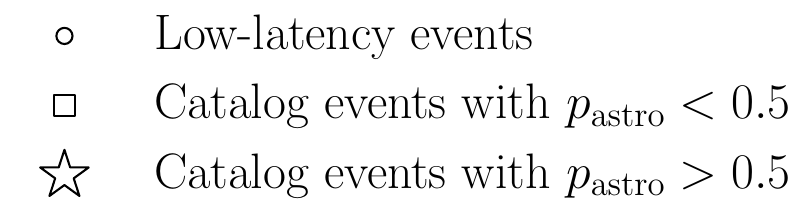}\\
    \includegraphics[width=1.05\columnwidth]{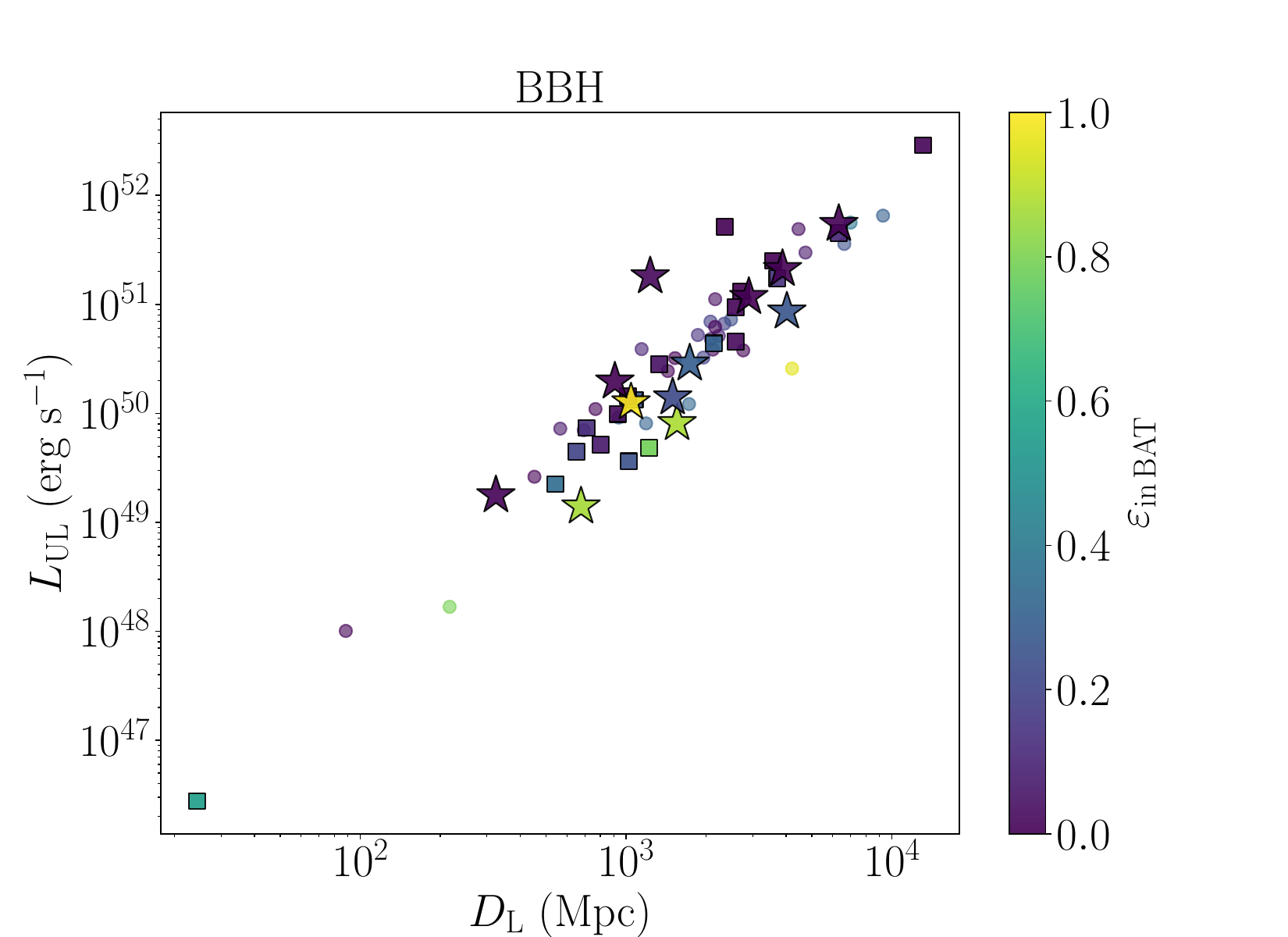}
    \includegraphics[width=1.05\columnwidth]{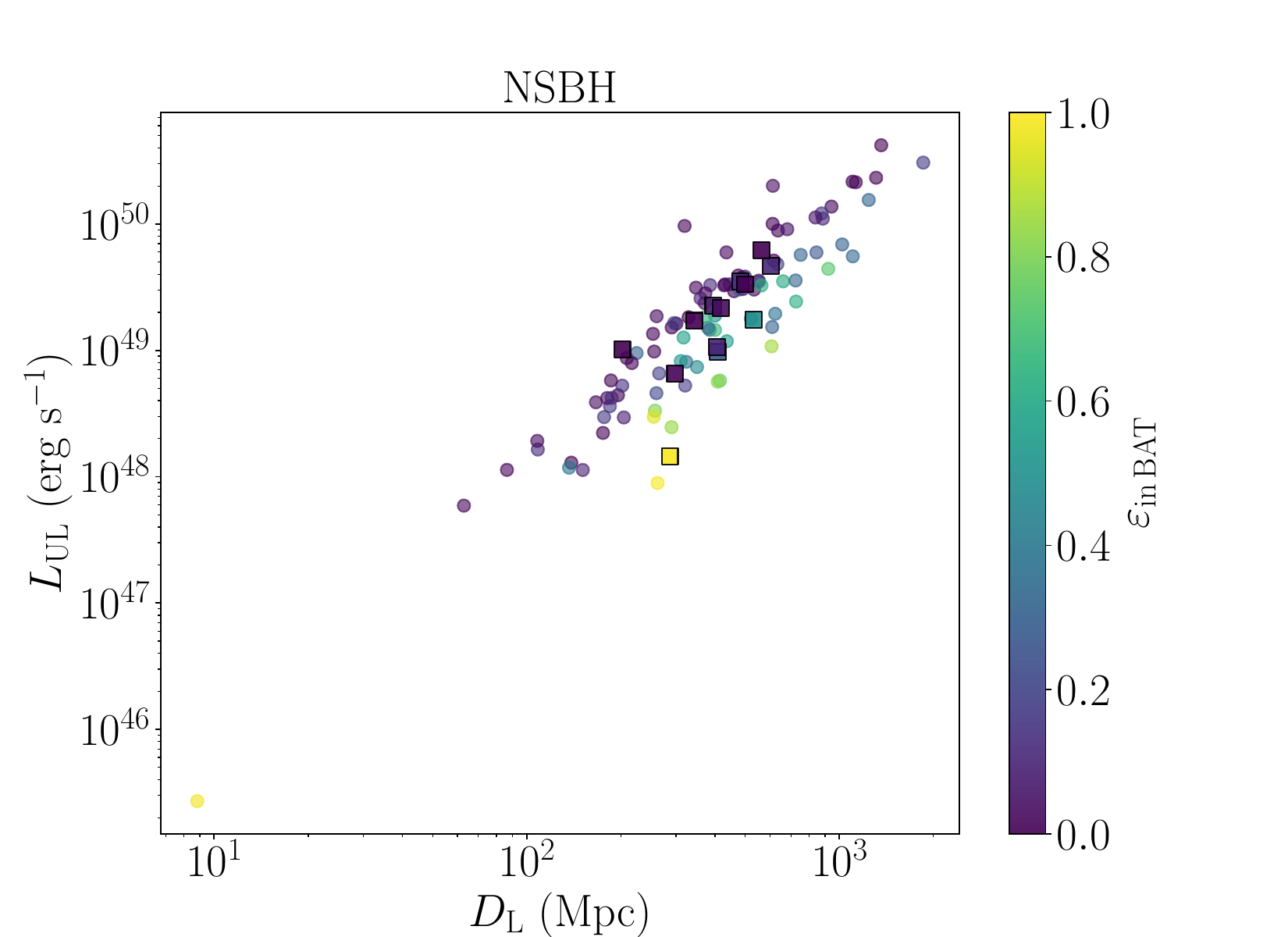}
    \includegraphics[width=1.05\columnwidth]{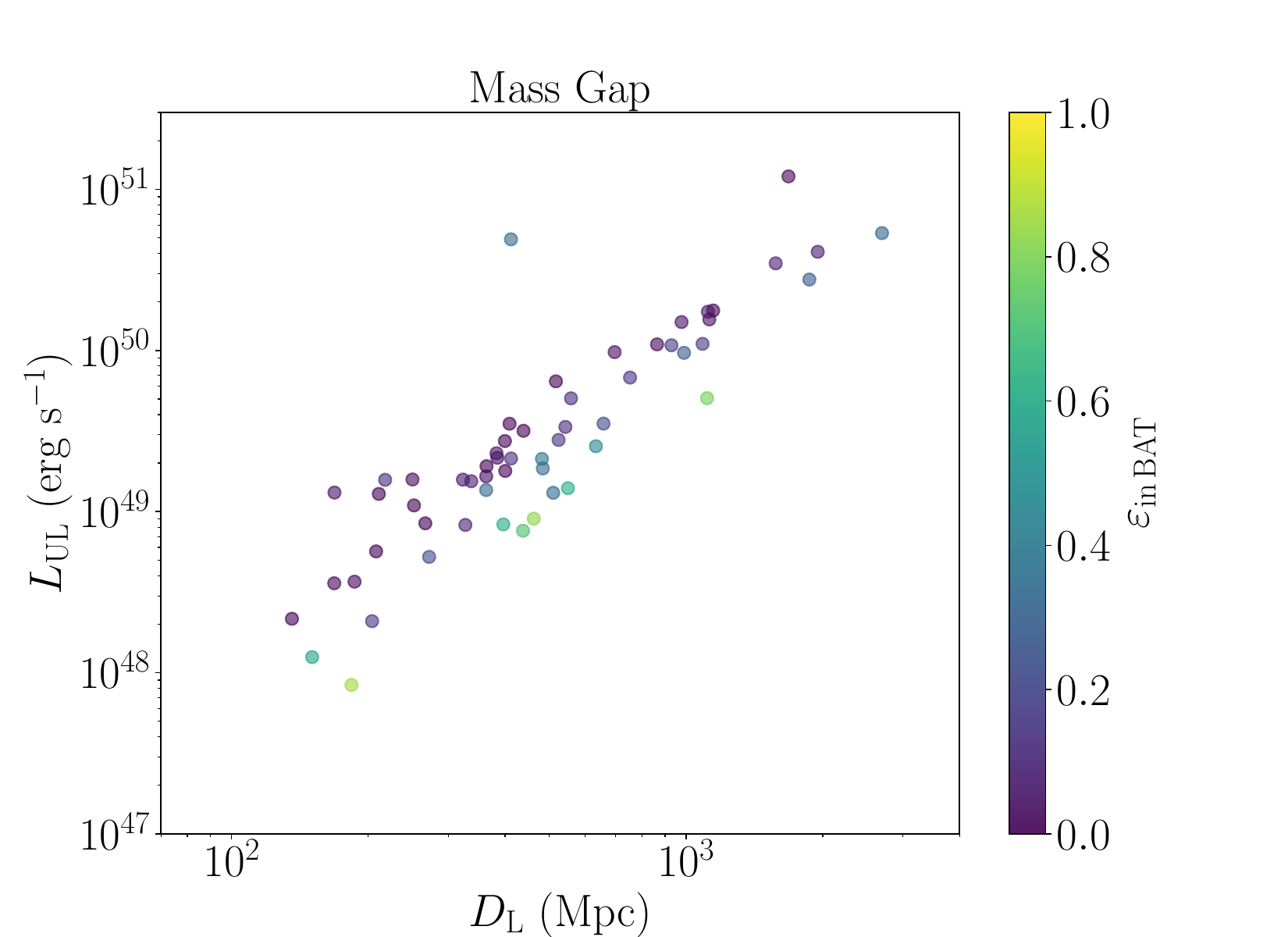}
    \includegraphics[width=1.05\columnwidth]{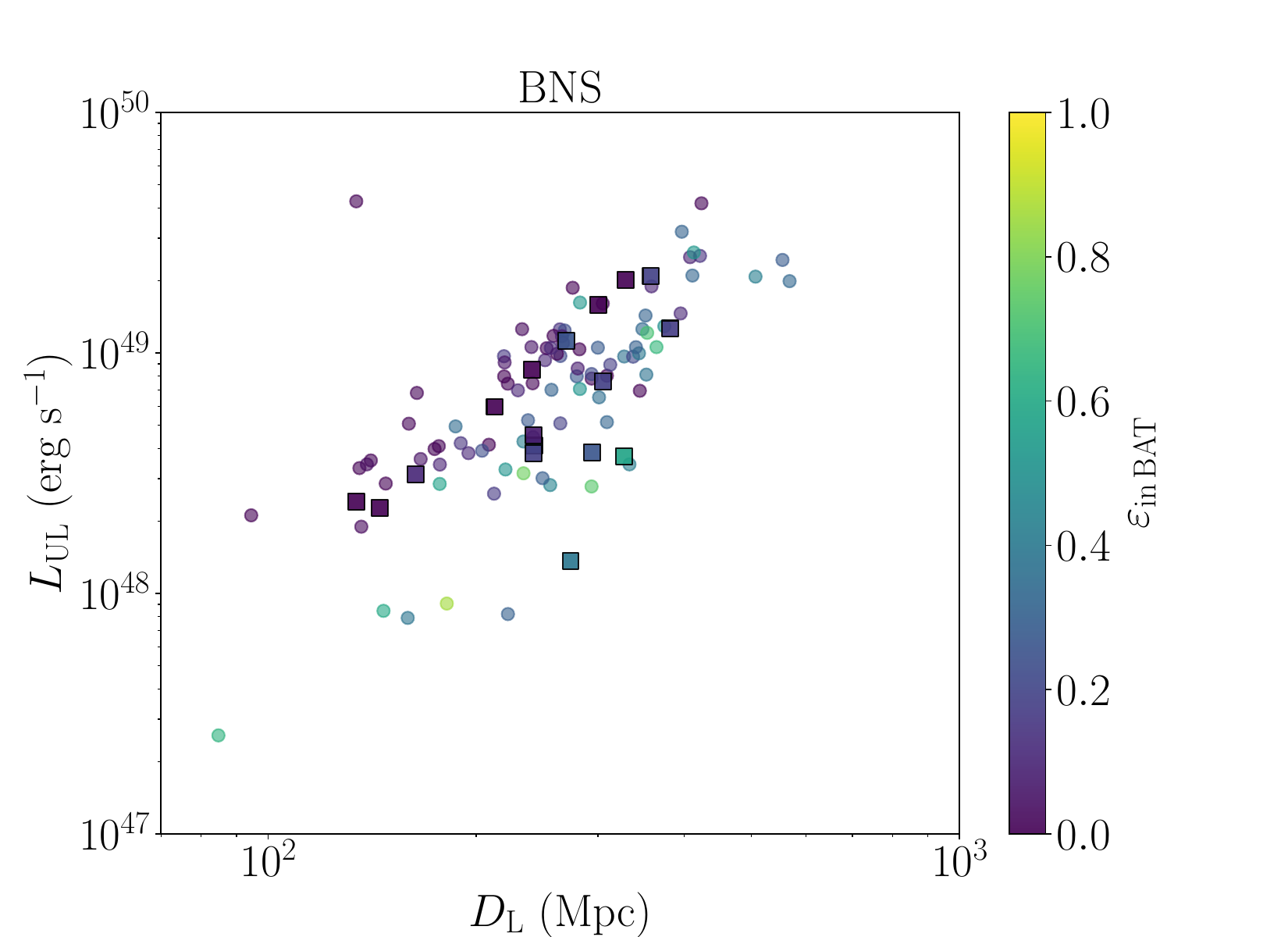}

    \caption{Upper limits on the luminosity computed in the rest frame 1~keV--10~MeV energy band, as a function of the mean luminosity distance extracted from the sky localization map of each GW candidate. The color bar indicates the quantity $\varepsilon_{\rm in \, BAT}$, namely the probability that the GW candidate is located inside the BAT coded FOV.  With different symbols, we distinguish the GW candidates received in low latency and the ones confirmed and included in the O3 catalog, separated in $p_{\rm astro}>0.5$ and $p_{\rm astro}<0.5$.}
    \label{fig:lum}
\end{figure*}

The luminosity upper limits as a function of the mean value of the luminosity distance is reported in Figure~\ref{fig:lum}. Similar to what was shown previously, we demarcate the various samples. Candidates with a low latency classification of Mass Gap were later re-distributed to other categories as part of post-processing, which is evident from the Mass Gap panel in Figure~\ref{fig:lum}.  As expected and as evident already from Figure~\ref{fig:gw_in}, we see a clear correlation between the luminosity upper limit and $\varepsilon_{\rm in \, BAT}$ in Figure~\ref{fig:lum}, indicating that the inferred constraints on the EM counterpart are more stringent when the GW probability integrated inside the BAT FOV is higher. Since $\phi_{\rm UL}$ is an upper limit and not a measure coming from a detection, Eq. (\ref{l_ul}) is an approximated method to convert $\phi_{\rm UL}$ in a luminosity upper limit, averaging over the $P(D_{\rm L})$ distribution provided by the GW analysis. In Appendix \ref{l_ul_app} we show a more accurate way to estimate the luminosity upper limit, but we find no relevant differences with respect to the method reported in this section. The Eq.~\eqref{l_ul} is used only to produce the plots of Fig.~\ref{fig:lum}, but this approximation is not used in Section \ref{science} to perform inference about the EM model parameters. Instead, in Section \ref{science} a reverse process is followed, namely the EM model is used to predict the luminosity, which is then convolved with $P(D_{\rm L})$ to obtain a probability distribution of the flux in the BAT energy band.

\subsection{Computation of joint FAR}\label{sec:far}

\begin{figure}
    \centering
    \includegraphics[width=1.1\linewidth]{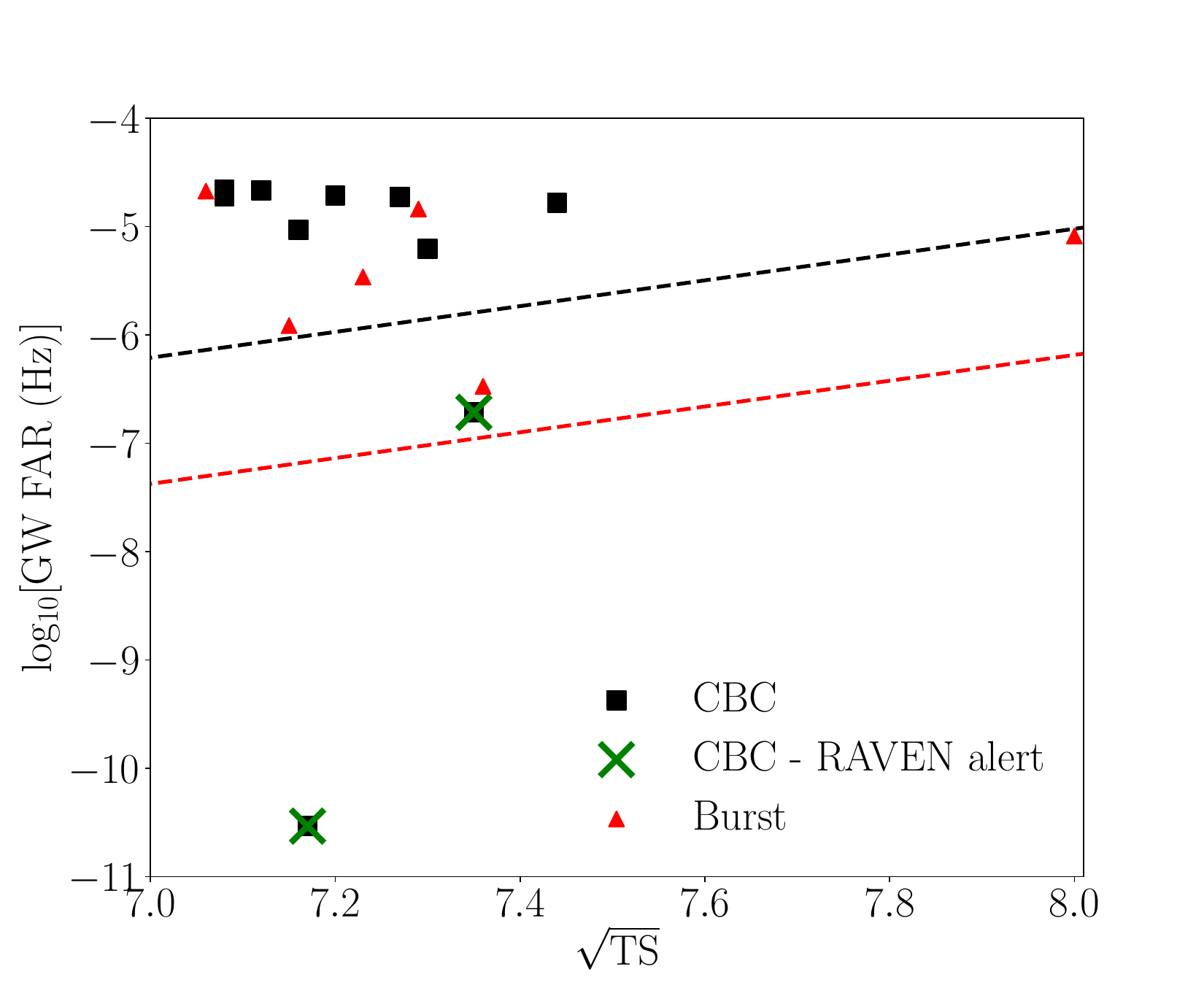}
    \caption{Distribution in the GW FAR--$\sqrt{\text{TS}}$ plane of all the triggers which passed the threshold FAR$_{\rm GRB}<10^{-3}$~Hz. Triggers in the regions below the black and red dashed lines (marked with a green cross) would have triggered the RAVEN alert system, for CBC and Burst events, respectively. The plot does not include all the triggers that have a $\sqrt{\text{TS}}>8$ which turned out to be spurious artifacts. }
    \label{fig:joint_far}
\end{figure}

To calculate the joint GW--BAT FARs for each GW trigger, we elaborate on the methods used to compute the individual FARs and subsequently combine them. To derive the sensitivity of the NITRATES search, time-tagged event data assembled from intervals corresponding to calibration runs and data from before and after known GRB signal times (total exposure time of $\sim$51~ks) were analyzed. The behavior of the background population and its associated FAR were then identified (see Section 7.6 and Fig.~16 in \citealt{NITRATES2022}).
We further compute the joint \textit{Swift}-BAT--GW FAR by combining the BAT FAR (calculated using the method described above) with the GW FAR. A targeted joint FAR threshold routine is constructed as part of the Rapid, on-source VOEvent Coincident Monitor (RAVEN;  \citealt{kindaRAVENref,RAVENthesis}), which combines the FARs obtained from GWs along with those from BAT and computes the joint temporal as well as the joint spatial FAR. We also specify details of the search pipeline used in the process, Burst or CBC. The joint FAR prescription as reported in the RAVEN documentation,\footnote{\url{https://lscsoft.docs.ligo.org/raven/joint_far.html}} is computed as 
\begin{equation}
    \text{FAR}_{\rm GRB+GW}=\frac{Z}{I_{\Omega}}\left[1+\ln ({\frac{Z_{\max}}{Z}})\right]
\end{equation}
where $Z$ is the joint ranking statistic given by,
\begin{equation}
Z=\text{FAR}_{\rm GW} \text{FAR}_{\rm GRB} \Delta t,
\end{equation}
\begin{equation}
Z_{\rm max}=\text{FAR}_{\rm GW,max} \text{FAR}_{\rm GRB,max} \Delta t,
\end{equation}
and we adopt $\Delta t=30$ s, $\text{FAR}_{\rm GW,max}=$ 2 day$^{-1}$ and $\text{FAR}_{\rm GRB,max}=10^{-3}$ Hz. $I_{\Omega}$ is an integral that quantifies the spatial overlap between the GW localization and the GRB localization \citep{Ashton2018}. Even if the search of subthreshold candidates in NITRATES is done in a temporal window $[t_0-20 \rm ~s, ~$$t_0+20 \rm ~s]$ around the trigger time $t_0$, for the RAVEN joint alert the adopted temporal window is $[t_0-10 \rm ~s, ~$$t_0+20 \rm ~s]$. Since none of the BAT candidates analyzed in this work has a confident estimation of the sky localization, we adopt a uniform posterior probability on the full sky for the EM candidate. Hence, by definition, we set $I_{\Omega}=1$. The candidate triggers a RAVEN alert when the $\text{FAR}_{\rm GRB+GW}\times N_t< \text{FAR}_{\rm max}$, with $N_t$ being the trials factor of the joint search and $\text{FAR}_{\rm max}=$ (1/30)~day$^{-1}$ for CBC events and $\text{FAR}_{\rm max}=$ 1~yr$^{-1}$ for Burst events. The trials factor corresponds to $N_t=S_{\rm GW}(S_{\rm GW}+1)$, where $S_{\rm GW}$ is the number of search GW pipelines, 4 for CBC events and 3 for Burst events \citep{Brandonthesis}. Since the GW pipelines are not fully independent, a realistic value of the trials factor is smaller than the one adopted here, therefore the RAVEN threshold can be considered as a conservative estimate for the significance of a joint detection.

We quote the derived joint FARs along with other trigger-specific details only for those triggers with a $\text{FAR}_{\rm GRB,max}<10^{-3}$ Hz in Table \ref{tab:jointfar}.
We find that, after rejecting false positives, 2 CBC events pass the joint FAR detection threshold to trigger a RAVEN alert. Specifically, S191110x ($\sqrt{\text{TS}}=7.2$) and S200108p ($\sqrt{\text{TS}}=7.4$) have a joint FAR of $3.02\times 10^{-4}$~yr$^{-1}$ and $21.3$~yr$^{-1}$, respectively. These values are obtained considering the GW FAR received with the low-latency alert. In the offline analysis of the GW candidates, neither S191110x or S200108p have been confirmed. We therefore conclude that, considering the offline joint analysis of GW and \textit{Swift}-BAT data, none of the candidates is eligible to claim a significant joint detection.

In Fig.~\ref{fig:joint_far} we report the location in the GW FAR - $\sqrt{\text{TS}}$ plane of all the candidates that pass the condition  $\text{FAR}_{\rm GW,max}<$2 day$^{-1}$ and $\text{FAR}_{\rm GRB,max}<10^{-3}$ Hz (i.e., $\sqrt{\text{TS}} \gtrsim 7$), to be considered for a potential joint alert. The astrophysical origin of all the candidates with $\sqrt{\text{TS}}>8$ has been rejected as discussed in Section \ref{sec:falsepos}, and therefore they are not reported in Fig.~\ref{fig:joint_far}. The dashed black and red lines mark the separation line for the event to pass the RAVEN alert threshold, for CBC and Burst candidates, respectively. Candidates below those lines would have triggered a RAVEN alert.

\section{Science discussion}

\label{science}

In this section, we describe how the upper limits derived from the joint subthreshold search can be used to infer constraints about possible EM emission from the GW candidates. Starting from a model of the EM emission, the luminosity in the BAT band can be estimated, whose value will depend on some internal parameters of the model $(\lambda_1, ..., \lambda_k)$. The goal is to explore the model parameter space and test if the estimated flux is in agreement with the upper limit constraints derived in this paper.

For this purpose, a knowledge of the distance of the GW candidate is needed, and the GW sky localization is used to extract the posterior distribution $P(D_{\rm L})$. Since only CBC events have such information, Burst events are not considered in this discussion. For the CBC events, we consider a phenomenological model which describes the probability distribution of the luminosity $L$ (in the $15$--$350$ keV rest-frame)
\begin{equation}
    P(L) = (1-f)\delta(L=0) + f \Pi(L).
\end{equation}
Here, the $f$ parameter is a proxy for the EM-bright nature of the event, i.e., given a CBC source described by a set of GW parameters $\Vec{\theta}_{\rm GW}$, $f(\Vec{\theta}_{\rm GW})$ corresponds to the probability that the EM luminosity of the source is non-zero. On the other hand, $\Pi(L)$ is the intrinsic luminosity function of the EM transient associated with the specific CBC class. In the case of BNS and NSBH candidates, the assumption on $\Pi(L)$ should be informed by our prior knowledge of the luminosity function of merger-driven GRBs. A detailed study of the impact of this work on our knowledge of merger-driven GRBs luminosity function will be reported in a follow-up paper.
In this section, instead, we focus only on the BBH class, for which no strong prior exists for $\Pi(L)$. For simplicity and in order to show the constraining power of our joint subthreshold search, we assume that the EM process associated with BBH, if present, produces a universal, viewing angle-independent luminosity $L_0$. 
Therefore, in the scenario specified above, we have $(\lambda_1,...,\lambda_k)=(f,L_0)$ and
\begin{equation}
  P(L) = (1-f)\delta(L=0) + f \delta(L-L_0) = P(L;f,L_0).  
\end{equation}

Once the model for the EM emission is specified, the probability distribution of the predicted flux is
\begin{equation}
    P(\phi)=(1-f)\delta(\phi=0)+f  P_{\text{EM}}(\phi),
\end{equation}
where $P_{\text{EM}}(\phi)=P(L_0/ 4\pi k D_{\text{L}}^2$) is the flux probability distribution in the assumption that the source is EM bright. Hence, for the $i$-th candidate, the probability that the predicted flux is below the estimated upper limit $\phi_{0,i}$ corresponds to
\begin{equation}
    P_i(\phi<\phi_{0,i})=(1-f)+f \int_0^{\phi_{0,i}} P_i(\phi) \mathrm{d} \phi,
\end{equation}
valid in the limit in which the GW candidate is assumed to be real.
Therefore, given a candidate GW with a probability of being astrophysical $p_{\text{astro},i}=\pi_i$, there are only three possibilities to have a non-detection in BAT:
\begin{enumerate}
    \item The source is not astrophysical, with a probability $1-\pi_i$; 
    \item The source is astrophysical, but it is occulted by the Earth, with a probability $\pi_i\varepsilon_{\Earth}$; 
    \item The source is astrophysical, it is not occulted by the Earth and the predicted flux by the EM model is below the BAT upper limit, with a probability $\pi_i(1-\varepsilon_{\Earth}) P_i(\phi<\phi_{0,i})$.
\end{enumerate}
This allows us to define a non-detection likelihood corresponding to
\begin{equation}
\label{li}
  \mathcal{L}_i=(1-\pi_i)+\pi_i[\varepsilon_{\Earth}+(1-\varepsilon_{\Earth})P_i(\phi<\phi_{0,i})]. 
\end{equation}
Additionally, there is also the possibility that the GW source is astrophysical, but mis-classified. As we discuss later, this has a negligible impact on our analysis. In the case of $L_0 \rightarrow 0$, $P_i(\phi<\phi_{0,i})\rightarrow1$, so $\mathcal{L}_i \rightarrow 1$. For very large values of $L_0$, instead, $P_i(\phi<\phi_{0,i})\rightarrow0$ and therefore $\mathcal{L}_i \rightarrow (1-\pi_i)+\pi_i\varepsilon_{\Earth}$. This last result shows how, even if the luminosity predicted by the model is exceedingly large, a non-detection can occur if the GW source is not real ($1-\pi_i$), or if it is real but occulted by the Earth ($\pi_i\varepsilon_{\Earth}$).
Since the analysis is focused only on BBH events, we consider only those candidates that have $p_{\rm BBH}>p_{\rm NSBH},p_{\rm BNS}$. By definition, $p_{\rm BBH}+p_{\rm NSBH}+p_{\rm BNS}=p_{\rm astro}$ and typically for the candidates classified as BBH we have that $p_{\rm BBH}\gg p_{\rm NSBH},p_{\rm BNS}$. The last condition allows us to consider Eq.~(\ref{li}) still valid if we replace $\pi_i$ with $p_{\text{BBH},i}$, since the contribution of $p_{\text{NSBH},i}$ and $p_{\text{BNS},i}$ to the non-detection probability is negligible. For all the excluded cases that have $p_{\rm BBH}<p_{\rm NSBH},p_{\rm BNS}$, we verified that $p_{\rm BBH}\ll 10^{-4}$, therefore their inclusion in the analysis would contribute minimally to our results.

Given the definition of Eq.~(\ref{li}), $\mathcal{L}_i$ indicates the probability, given a set of $(\lambda_1, ..., \lambda_k)$ EM parameters, that the BAT upper limit is not violated, taking into account the possible non-astrophysical origin of the candidate and also the probability that, even if astrophysical, the source is occulted by the Earth and therefore not detectable by \textit{Swift}.
Having a collection of $E_1,...,E_N$ GW candidates, the posterior distribution of the model parameters can be obtained following the Bayes theorem
\begin{gather}
P(L_0,f|E_1,...,E_N) \\ \notag 
=\prod_{i=1}^N \mathcal{L}_i \pi(L_0)\pi(f)\Big/\int \prod_{i=1}^N \mathcal{L}_i \pi(L_0)\pi(f) \mathrm{d}L_0 \mathrm{d}f\label{eq1},
\end{gather}
where $\pi(L_0)$ and $\pi(f)$ are the prior distributions of $L_0$ and $f$. We assume a log-uniform prior for both $L_0$ and $f$ in the respective intervals $46<\log_{10}[L_0(\rm erg~s^{-1})]<53$ and $-3<\log_{10}(f)<0$. The choice of the prior boundaries are poorly  informed by theoretical expectations, which are still affected by large uncertainties. Instead, the priors are chosen on the basis of the typical range of upper limit luminosity derived in this work for BBH events and the total number of candidates considered in this analysis. The constraints reported in the following may strongly depend on the choice of the prior boundaries. Therefore, the final goal of this simulation, more than deriving strong limits on the putative EM model, is to show the predictive power of the present analysis in the context of model inference and how this analysis can improve with the addition of more GW events in the future. 

In the specific case of our simulation, we consider all the GW candidates released in GWTC-3 \citep{GWTC3-2023PhRvX}, including both the above threshold ($p_{\rm astro}>0.5$) and the subthreshold ($p_{\rm astro}<0.5$) candidates. For the latter, we emphasize that the classification of the CBC candidate as BBH merger is valid under the condition that the subthreshold GW event is of astrophysical origin. The simulation described in this section is set up in such a way that this assumption is taken into account for the final constraints of the physical parameters. All the low-latency candidates not confirmed by the offline analysis are not included in the simulation. The considered BBH sample with full NITRATES results and available flux upper limits consists of $32$ events, 12 of which with $p_{\rm astro}>0.5$.

\begin{figure}
    \centering
    \includegraphics[width=1\linewidth]{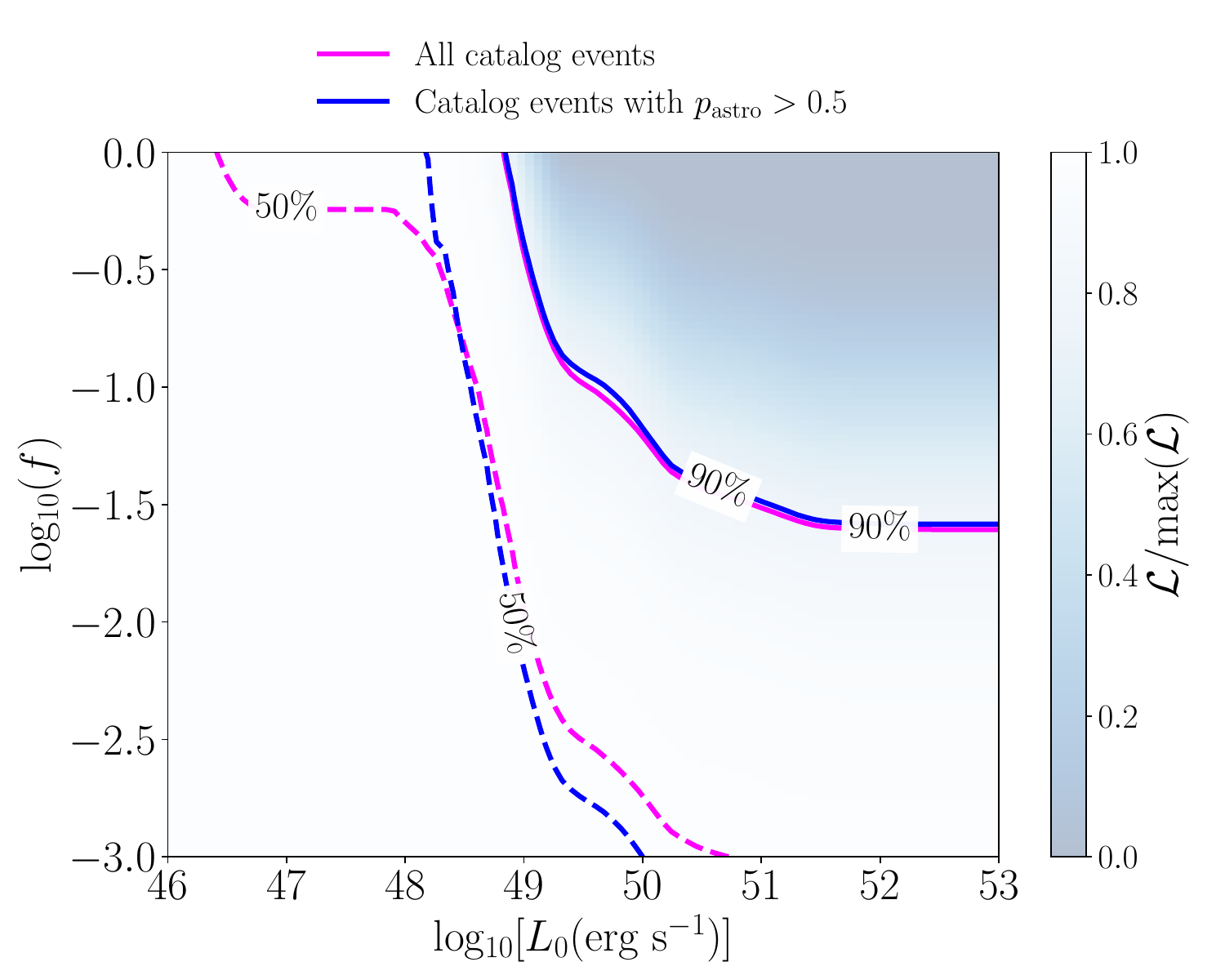}
    \caption{Constraints on the two parameters $L_0$ and $f$ of the model for the putative EM counterpart of BBH mergers. The color map reports the likelihood $\mathcal{L}$, for the full analysis including all the O3 catalog events with $p_{\rm astro}>0.5$. $L_0$ is in units of erg s$^{-1}$. The thick blue solid and dashed contours indicate the exclusion regions in the $[L_0,f]$ plane at 90$\%$ and 50$\%$ credibility levels, respectively. The magenta solid and dashed lines report the same contours, but for an analysis that includes all O3 catalog events, with no cut in $p_{\rm astro}$.}
    \label{fig:BBH_NH}
\end{figure}

\begin{figure}
    \centering
    \includegraphics[width=1\linewidth]{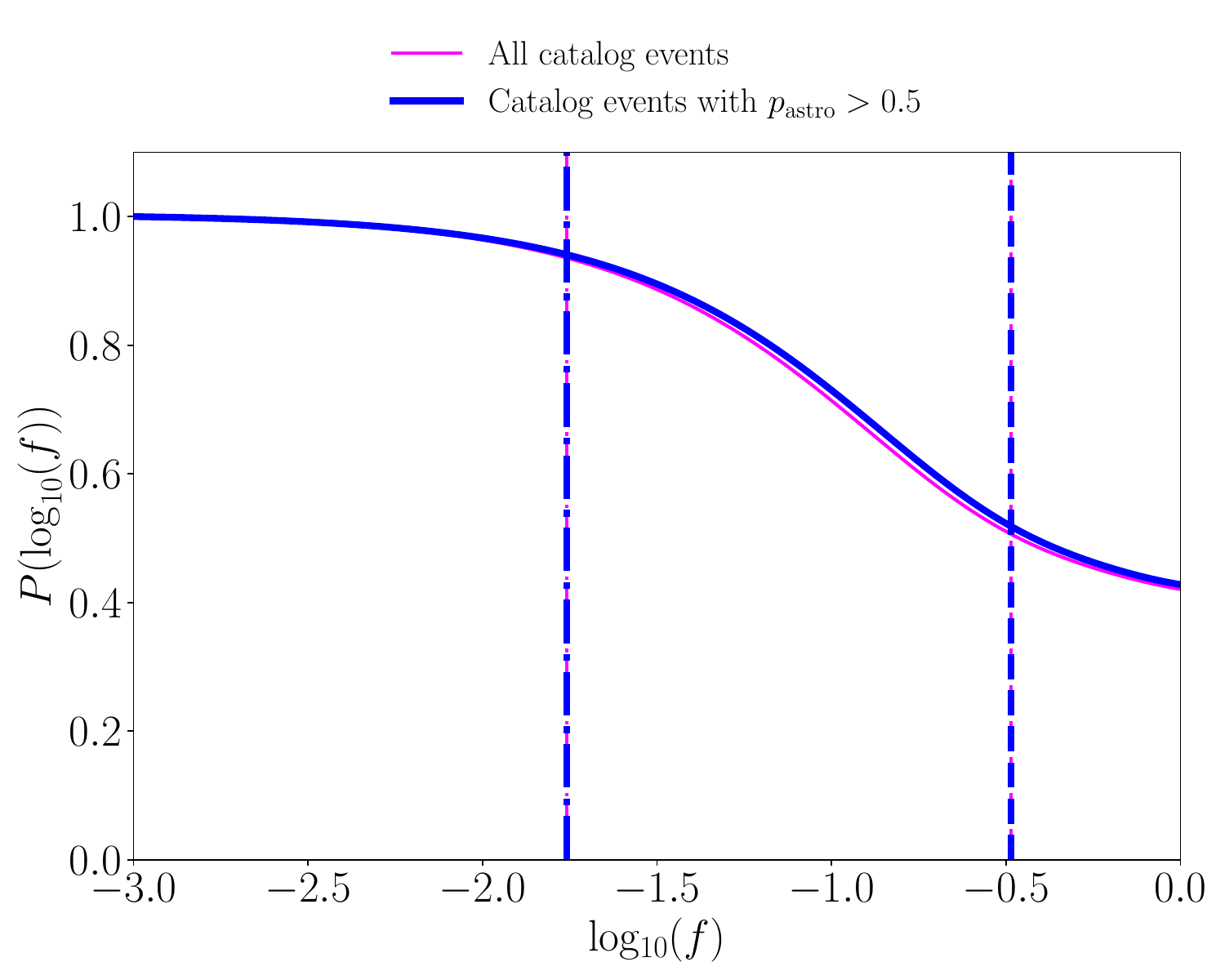}
    \caption{Posterior distribution of $\log_{10}(f)$, including the 50$\%$ and 90$\%$ upper limits with dot-dashed and dashed lines, respectively. The function is derived from Fig.~\ref{fig:BBH_NH}, marginalizing over $L_0$.}
    \label{fig:postf}
\end{figure}

\begin{figure}
    \centering
    \includegraphics[width=1\linewidth]{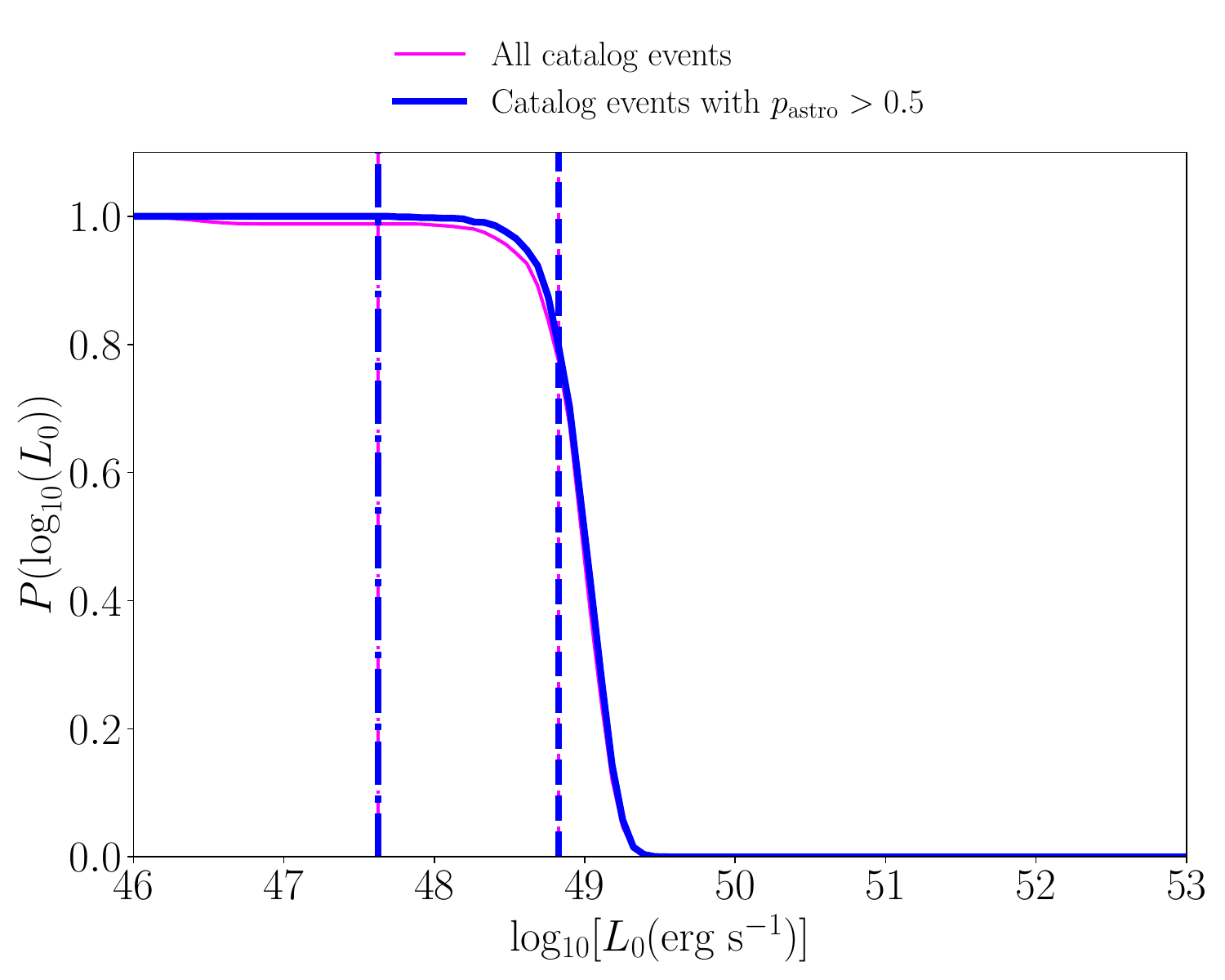}
    \caption{Posterior distribution of $\log_{10}(L_0)$, including the 50$\%$ and 90$\%$ upper limits with the dot-dashed and dashed lines, respectively. The function is derived from Fig.~\ref{fig:BBH_NH}, marginalizing over $f$.}
    
    \label{fig:postl0}
\end{figure}

In order to compute numerically the functional behavior of the likelihood, we set up a simulation to evaluate $P(L_{0},f|E_1,...,E_N)$ in the full $[L_{0},f]$ plane defined by the prior boundaries. The details of the simulation setup are reported in Appendix \ref{sim_set}. The results of the simulation for the sample of above threshold BBH candidates are reported in Fig.~\ref{fig:BBH_NH}, where the color map indicates the value of $\mathcal{L}=\prod \mathcal{L}_i$, normalized by the maximum $\max{(\mathcal{L})}$ over the full domain. The contour levels defining the 50$\%$ and 90$\%$ exclusion regions are reported as well. For comparison, in Fig.~\ref{fig:BBH_NH} we include also the same contour levels obtained with an analysis that considers all the BBH candidates without imposing any cut on the $p_{\rm astro}$. The level of constraining power of the analysis can be quantified by defining the fraction of the full parameter space excluded with a credibility level $\eta$, corresponding to:
\begin{equation}
 R_{\eta}=I_{\eta}/I_{\rm tot},
\end{equation}
where $I_{\text{tot}}=\int \text{d}f \text{d}L_0$, being the integral extended to the full parameters domain, and with
\begin{equation}
    I_{\eta}=\int_{S} \mathrm{d}f \mathrm{d}L_0,
\end{equation}
corresponding to the dimension of the region $S$ of the parameter space excluded with a credibility level $\eta$. The analysis performed using only BBH with $p_{\rm astro}>0.5$  gives a $R_{90\%}=26.0\%$, while the analysis performed with the inclusion of subthreshold BBH candidates gives $R_{90\%}=26.5\%$. This result indicates that with the inclusion of GW subthreshold events, the analysis allows us to exclude a slightly larger portion of the parameter space, with respect to an analysis carried out using only events with high $p_{\rm astro}$. 

Figs.~\ref{fig:postf} and \ref{fig:postl0} report the posterior distribution of $P(f)$ and $P(L_0)$, respectively, obtained as:
\begin{equation}
    P(f)=\int P(L_0,f|E_1,...,E_N)\mathrm{d}L_0
\end{equation}
and
\begin{equation}
P(L_0)=\int P(L_0,f|E_1,...,E_N)\mathrm{d}f.
\end{equation}
Both $P(L_0)$ and $P(f)$ are normalized such that $\max[P(L_0)]=\max[P(f)]=1$. The posterior is reported in magenta and blue for both samples, with and without cut in $p_{\rm astro}$, respectively. The $50\%$ and $90\%$ upper limits are reported as well. From the shape of the posteriors, it is evident that both the $L_0$ and $f$ posteriors change slightly if no cut in $p_{\rm astro}$ is applied.
For the sample with $p_{\rm astro}>0.5$, the $50\%$ and $90\%$ upper limits for $f$ are $\log_{10}(f_{50\%})=-1.76$ and $\log_{10}(f_{90\%})=-0.48$, while for $L_0$ are $\log_{10}[L_{0,50\%}(\text{erg s}^{-1})]=47.6$ and $\log_{10}[L_{0,90\%}(\text{erg s}^{-1})]=48.8$, respectively.

In the limit of a collection of triggers which correspond only to non-astrophysical events, i.e., all with $\pi_i=0$, the likelihood is constant in the full parameter space, not allowing to infer any constraints on the EM model parameters. On the other hand, if we increase the fraction of confident GW events and we keep fixed the total number $N$, their distance distribution $P(D_{\text{L}})$ and the derived upper limits,
then we obtain that $\mathcal{L}$ decreases accordingly. This implies that increasing the number of events with $\pi_i$ close to 1, the overall exclusion region in the $(\lambda_1, ..., \lambda_k)$ parameter space increases as well. This demonstrates that with the collection of more data, in the limit of a GW detector horizon constant in time, this method allows us to improve incrementally our constraints on the EM models of CBC events. Although, realistically the GW detection horizon will increase with time \citep{2020LRR....23....3A}, implying an overall increase of the median values of $D_{\rm L}$ of the candidate events. Such an effect increases in turn the values of the luminosity upper limits, increasing as well the values of $P_i(\phi<\phi_0)$ and hence of $\mathcal{L}$. This effect tends to decrease the dimension of the exclusion region of the $(\lambda_1, ..., \lambda_k)$ parameter space. Overall, the final outcome of the inclusion of additional GW data, in terms of the constraining power of this analysis, will depend on the simultaneous combined effect of increasing the number of confident events and of increasing the detection horizon. 

\begin{figure}
    \centering
    \includegraphics[width=1\linewidth]{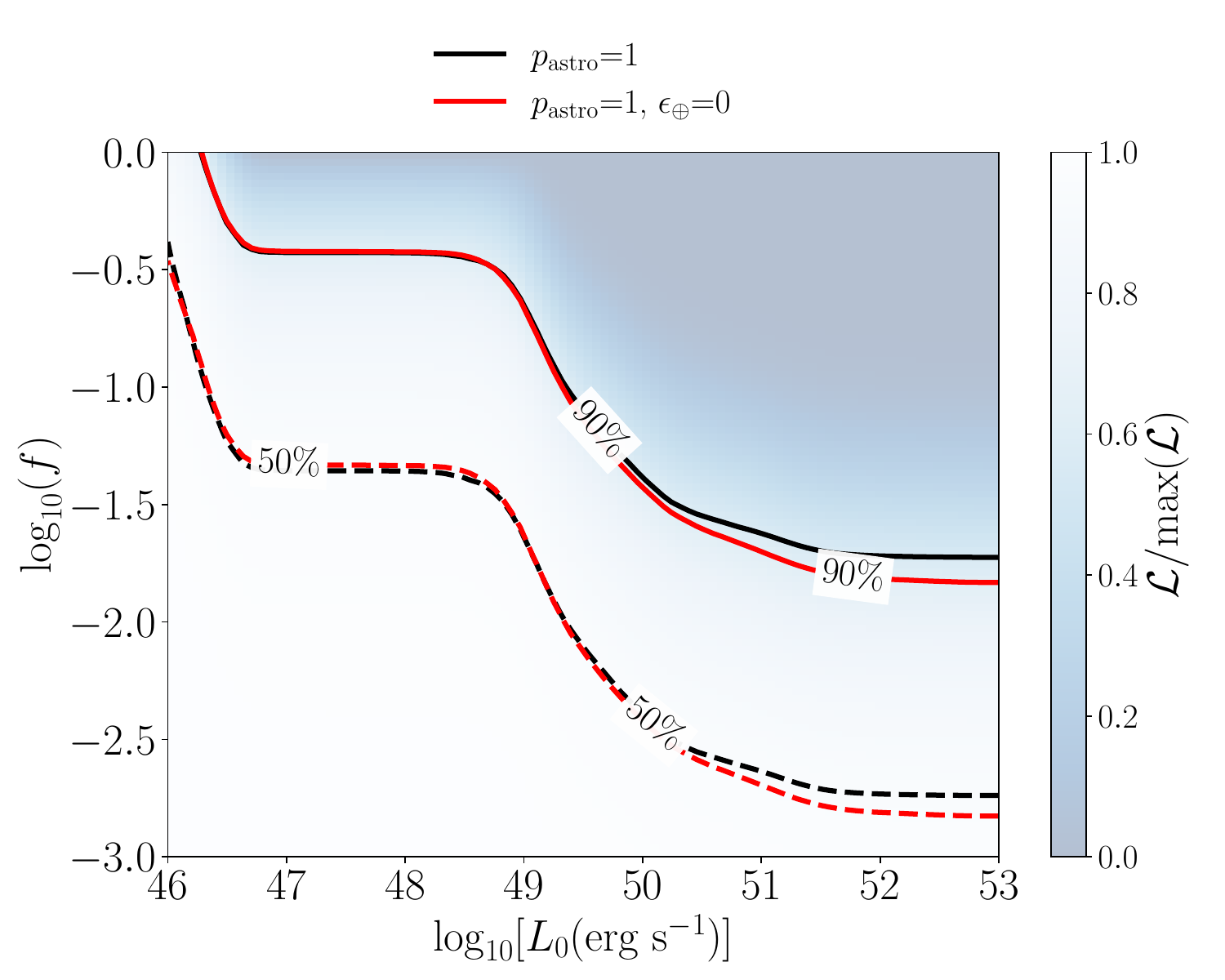}
    \caption{Same as Fig.~\ref{fig:BBH_NH}, but simulating all the O3 catalog candidates with an associated $p_{\rm astro}=\pi_i=1$. The black dashed and solid lines identify the $50\%$ and $90\%$ exclusion regions, respectively. The red dashed and solid lines have the same meaning, but derived imposing both $\pi_i=1$ and $\varepsilon_{\Earth}=0$.}
    \label{fig:BBH_allreal}
\end{figure}

In order to show how the inclusion of more significant GW candidates can improve the constraining power of the present analysis, we carried out the following simulation. We repeated the same procedure adopted to produce the exclusion regions of Fig.~\ref{fig:BBH_NH}, but replacing the real $\pi_i$ with $\pi_i=1$ for all the confirmed BBH candidates, hence imposing that they are all significant events. All the values of $\phi_{\rm UL}$, $\varepsilon_{\Earth}$ and $P(D_{\rm L})$ of each candidate are left unchanged. The resulting 50$\%$ and 90$\%$ exclusion regions are reported in Fig.~\ref{fig:BBH_allreal}, with black dashed and solid lines, respectively. The fraction of the 90$\%$ excluded region increases to a value of $R_{90\%}=34.2\%$, clearly demonstrating that, even if the BAT flux upper limit are the same, the increase of confidence about the astrophysical nature of the GW improves our final constraints on the model parameter space. Furthermore, Fig.~\ref{fig:BBH_allreal} reports also the 50$\%$ and 90$\%$ exclusion regions (with red dashed and solid lines, respectively), obtained as before, but imposing both $\pi_i=1$ and $\varepsilon_{\Earth}=0$ for each candidate. This combination corresponds to simulate all real BBH candidates, whose sky localization does not overlap with the sky region covered by the Earth. In this case $R_{90\%}=35.4\%$, showing that the fraction of GW sky posterior occulted by Earth has a slight impact on our final results.

\section{Conclusions}

In this work we report the systematic search of signals jointly detected by the LIGO--Virgo interferometers and the \textit{Swift}-BAT telescope, during the third LVK observing run. Thanks to the prompt availability of BAT data using GUANO and the sensitive targeted search capabilities of the NITRATES pipeline, we conducted deep follow-up searches for EM signals on a sample of 636 GW triggers. The search results did not yield any confident joint detection, allowing us to derive upper limits in the 15--350~keV band. We provide comprehensive details on all analyzed GW triggers along with their NITRATES search statistics. This information can be valuable for calibrating and comparing with other offline targeted search pipelines that are currently operational or may be developed in the future.

In the specific case of the BBH class, the BAT flux upper limits have been used to perform a stacking analysis and to derive constraints on the possible nature of an associated EM emission. As illustrated in Section \ref{science}, the presence of several BBH candidates with large values of $p_{\rm astro}$ in our sample, enhances our ability to better constrain the parameter space for EM emission from BH mergers, with minimal assumptions on our prior knowledge of the nature of the emission. The prospect of detecting EM emission from BBH mergers has been debated and discussed in detail in recent times. Particularly, the GBM trigger that accompanied the first BBH merger event, GW150914, has served as a case study to test possible association and potential implications \citep{AbbottGW1509142016PhRvL,Connaughton2016ApJ,Goldstein2016arXiv}. Though not likely, there are a number of physical models that have been proposed that could give rise to detectable emission in the gamma-ray band. A summary of the various different models has been discussed in \citet{Fletcher2023arxiv} and \citet{Veres2019ApJ}. The models involve parameters pertaining to potential remnant accretion effects, magnetic field strength, black hole charge and spin, among others (e.g.,  \citealt{Loeb2016ApJ,Dai2017,Woosley2016ApJ,Zhang2016ApJ}). The method described in Section \ref{science} can be easily extended to any of these models, provided that the luminosity function $\Pi(L)$ of the putative BBH EM emission is known. Additionally, effects possibly related to the viewing angle dependency of the EM emission can be easily included in this approach. Regarding CBCs containing at least one NS (BNS and NSBH), it was not possible to conduct a similar stacking analysis as the one described for BBH in Section \ref{science}, due to the paucity of such events with a large enough value of $p_{\rm astro}$. Further observations, including the fourth LVK observing run (O4), could lead to the collection of a larger number of BNS and NSBH candidates with moderate values of $p_{\rm astro}$, giving the possibility to repeat the analysis performed in this paper and to derive informative constraints on the EM emission of these classes and the properties of the associated GRB populations. Data products associated with the present analysis are reported in a separate data release.\footnote{\url{https://doi.org/10.5281/zenodo.10600302}}

O4 commenced on the 24th of May, 2023. The number of significant detections is expected to increase by several times during the entire duration of O4 \citep{Abbott-propsects2018LRR,Petrov2022ApJ}. Targeted search results using the GUANO-NITRATES infrastructure are publicly available in real-time\footnote{\url{https://guano.swift.psu.edu}}. In the case of non-detection of an EM counterpart, the GUANO team reports the 15--350~keV flux upper limit for all the GW triggers classified as \textit{significant}, via GCN Circulars. Additional enhancements to the likelihood search code have reduced the search latency by a factor of 2, with respect to O3.

Thanks to its sensitivity in the hard X-ray band and the possibility to localize EM transients down to a precision of an arcminute, \textit{Swift} represents one of the main discovery machines for the detection of EM counterparts of GW transients. This paper shows how the GUANO infrastructure has a deep impact on the multi-messenger science case, in particular for optimally exploiting the sensitivity of the \textit{Swift}-BAT instrument for the detection of EM counterparts of CBCs detected by the LVK Collaboration. The deep subthreshold search enabled by the NITRATES pipeline sensibly increases the detection horizon of \textit{Swift}, giving the chance to detect transients also outside the BAT FOV and allowing us to possibly detect faint X-ray/gamma-ray transients associated to relativistic jets observed off-axis, as in the case of GW170817. In the case of a confident joint \textit{Swift}-GW detection, the GUANO team will promptly disseminate all the information about the EM candidate via GCN Circulars, providing an estimate of the sky localization when available. Moreover, also in the case of non-detection, this paper shows how the upper limits derived from the NITRATES analysis can be combined to have the most sensitive constraints on the EM emission from all the CBC classes. The cumulative collection of non-detection will gradually improve our knowledge of the EM nature of CBCs.

\section*{Acknowledgments}
Gayathri Raman, Samuele Ronchini, and Jamie Kennea acknowledge the support of NASA grants 80NSSC19K0408 and 80NSSC22K1498 awarded as part of the NASA Neil Gehrels Swift Observatory Guest Investigator program. Jamie Kennea and James Delaunay acknowledge the support of NASA contract NAS5-0136.

This material is based upon work supported by NSF’s LIGO Laboratory which is a major facility
fully funded by the National Science Foundation.
The authors also gratefully acknowledge the support of
the Science and Technology Facilities Council (STFC) of the
United Kingdom, the Max-Planck-Society (MPS), and the State of
Niedersachsen/Germany for support of the construction of Advanced LIGO 
and construction and operation of the GEO\,600 detector. 
Additional support for Advanced LIGO was provided by the Australian Research Council.
The authors gratefully acknowledge the Italian Istituto Nazionale di Fisica Nucleare (INFN),  
the French Centre National de la Recherche Scientifique (CNRS) and
the Netherlands Organization for Scientific Research (NWO), 
for the construction and operation of the Virgo detector
and the creation and support  of the EGO consortium. 
The authors also gratefully acknowledge research support from these agencies as well as by 
the Council of Scientific and Industrial Research of India, 
the Department of Science and Technology, India,
the Science \& Engineering Research Board (SERB), India,
the Ministry of Human Resource Development, India,
the Spanish Agencia Estatal de Investigaci\'on (AEI),
the Spanish Ministerio de Ciencia, Innovaci\'on y Universidades,
the European Union NextGenerationEU/PRTR (PRTR-C17.I1),
the ICSC - CentroNazionale di Ricerca in High Performance Computing, Big Data
and Quantum Computing, funded by the European Union NextGenerationEU,
the Comunitat Auton\`oma de les Illes Balears through the Direcci\'o General de Recerca, Innovaci\'o i Transformaci\'o Digital with funds from the Tourist Stay Tax Law ITS 2017-006,
the Conselleria d'Economia, Hisenda i Innovaci\'o, the FEDER Operational Program 2021-2027 of the Balearic Islands,
the Conselleria d'Innovaci\'o, Universitats, Ci\`encia i Societat Digital de la Generalitat Valenciana and
the CERCA Programme Generalitat de Catalunya, Spain,
the National Science Centre of Poland and the European Union – European Regional Development Fund; Foundation for Polish Science (FNP),
the Polish Ministry of Science and Higher Education,
the Swiss National Science Foundation (SNSF),
the Russian Science Foundation,
the European Commission,
the European Social Funds (ESF),
the European Regional Development Funds (ERDF),
the Royal Society, 
the Scottish Funding Council, 
the Scottish Universities Physics Alliance, 
the Hungarian Scientific Research Fund (OTKA),
the French Lyon Institute of Origins (LIO),
the Belgian Fonds de la Recherche Scientifique (FRS-FNRS), 
Actions de Recherche Concertées (ARC) and
Fonds Wetenschappelijk Onderzoek – Vlaanderen (FWO), Belgium,
the Paris \^{I}le-de-France Region, 
the National Research, Development and Innovation Office Hungary (NKFIH), 
the National Research Foundation of Korea,
the Natural Science and Engineering Research Council Canada,
Canadian Foundation for Innovation (CFI),
the Brazilian Ministry of Science, Technology, and Innovations,
the International Center for Theoretical Physics South American Institute for Fundamental Research (ICTP-SAIFR), 
the Research Grants Council of Hong Kong,
the National Natural Science Foundation of China (NSFC),
the Leverhulme Trust, 
the Research Corporation,
the National Science and Technology Council (NSTC), Taiwan,
the United States Department of Energy,
and
the Kavli Foundation.
The authors gratefully acknowledge the support of the NSF, STFC, INFN and CNRS for provision of computational resources.

This work was supported by MEXT, JSPS Leading-edge Research Infrastructure Program, JSPS Grant-in-Aid for Specially Promoted Research 26000005, JSPS Grant-in-Aid for Scientific Research on Innovative Areas 2905: JP17H06358, JP17H06361 and JP17H06364, JSPS Core-to-Core Program A. Advanced Research Networks, JSPS Grant-in-Aid for Scientific Research (S) 17H06133 and 20H05639 , JSPS Grant-in-Aid for Transformative Research Areas (A) 20A203: JP20H05854, the joint research program of the Institute for Cosmic Ray Research, University of Tokyo, National Research Foundation (NRF), Computing Infrastructure Project of Global Science experimental Data hub Center (GSDC) at KISTI, Korea Astronomy and Space Science Institute (KASI), and Ministry of Science and ICT (MSIT) in Korea, Academia Sinica (AS), AS Grid Center (ASGC) and the National Science and Technology Council (NSTC) in Taiwan under grants including the Rising Star Program and Science Vanguard Research Program, Advanced Technology Center (ATC) of NAOJ, and Mechanical Engineering Center of KEK.

Additional acknowledgements for support of individual authors may be found in the following document: \\
\texttt{https://dcc.ligo.org/LIGO-M2300033/public}.
For the purpose of open access, the authors have applied a Creative Commons Attribution (CC BY)
license to any Author Accepted Manuscript version arising.
We request that citations to this article use 'A. G. Abac {\it et al.} (LIGO-Virgo-KAGRA Collaboration), ...' or similar phrasing, depending on journal convention.

Matplotlib~\citep{MatplotlibHunter},  SEABORN~\citep{SeabornWaskom2021},   
NumPy~\citep{NumPy2020Natur} and SciPy~\citep{Virtanen2020NatMe..17..261V} were used in the preparation of the manuscript.

\clearpage

\bibliographystyle{mnras}
\bibliography{bibtex}

\newpage
\appendix

\section{luminosity upper limit}
\label{l_ul_app}
A more accurate method to derive the luminosity upper limit should be based on the knowledge of $P(D_{\rm L})$ and $P(\phi)$, where $\phi$ is the flux measured in the BAT energy band. Having only an upper limit, $P(\phi)$ can be approximated as
\begin{equation}
    P(\phi)\propto
    \begin{cases}
        \Pi(\phi), & \phi<\phi_{\text{UL}} \\
        0, & \phi>\phi_{\text{UL}}
    \end{cases},
\end{equation}
where $\Pi(\phi)$ is our prior distribution for the flux. Using the conversion from flux to luminosity $L=4\pi D_{\rm L}^2 \phi$, the probability distribution of the luminosity can be computed as
\begin{equation}
    P(L) = P(4 \pi D_{\rm L}^2 \phi) \propto  \int  \frac{1}{\phi} P_{\phi}(\phi) P_{D_{\rm L}^2}\left(\dfrac{L}{4\pi\phi}\right) \rm d \phi,
\end{equation}
where $P_{\phi}$ is the flux probability distribution and $P_{D_{\rm L}^2}$ is the probability distribution of $D_{\rm L}^2$. In the conversion from flux to luminosity, the k-correction has been omitted, since it introduces a mild dependence on the redshift, which is not relevant for the purposes of this section. The 5$\sigma$ luminosity upper limit $L_{\text{UL}}$ can be found imposing that
\begin{equation}
   \int_0 ^{L_{\text{UL}}} P(L) \text{d} L = 1-\varepsilon_{5\sigma},
\end{equation}
with $\varepsilon_{5\sigma}=3\times 10^{-7}$. The value of $L_{\text{UL}}$ has been computed adopting two different assumptions for the flux prior, corresponding to $\Pi(\phi)\propto $ const. and $\Pi(\phi)\propto \phi^{-3/2}$, with the latter being inspired by the usual trend followed by GRBs (e.g., \citealt{om2023A&A...680A..45S}). In both cases, we find that $L_{\text{UL}}\sim 3 \times 4 \pi \langle D_{\rm 
L}^2 \rangle \phi_{\text{UL}}$.

\section{Simulation setup}
\label{sim_set}
In this section we specify the details of the simulation used to compute numerically the $\mathcal{L}(L_{0},f|E_1,...,E_N)$ function. For each simulated GW candidate, the single $\mathcal{L}_i$ is computed for each pairs of values $(L_{0,n},f_m)$. The flux predicted by the EM model is predicted injecting 1000 sources whose luminosity distance is distributed according to $P(D_{\rm L})$, derived from the GW localization. The probability $P_i(\phi<\phi_{0,i})$ is derived computing the fraction of cases that have a flux below the sky-averaged BAT upper limit, defined by Eq.~\eqref{phi_ul}. The computation of $\mathcal{L}$ is performed on a $100\times100$ grid of $(L_{0,n},f_m)$. Once the previous steps are performed for all the GW candidates, the final combined likelihood is computed as
\begin{equation}
    \mathcal{L}(L_{0},f|E_1,...,E_N) = \prod_i \mathcal{L}_i(L_{0},f).
\end{equation}
In order to produce the credibility contours in the $[L_{0,n},f_m]$ plane, we adopt the following steps: 
\begin{enumerate}
    \item $\mathcal{L}$ is normalized such that
    \begin{equation}
    \sum_{n,m}\mathcal{L}(L_{0,n},f_m)=1.
    \end{equation}
    \item A one-dimensional array $\mathcal{L}[x_n]$ is created  flattening the  two-dimensional grid $\mathcal{L}(L_{0,n},f_m)$, then $\mathcal{L}[x_n]$ is sorted in ascending order.
    \item We find the element $[\hat{L}_{0},\hat{f}]=[x_{n^{*}}]$ such that
    \begin{equation}
    \sum_{n=0}^{n^{*}} \mathcal{L}[x_n]=\lambda,
    \end{equation}
    where $\lambda$ is the credibility level of the contour.
    \item The contour is drawn imposing $\mathcal{L}=\mathcal{L}(\hat{L}_{0},\hat{f})$.
\end{enumerate}

 \section{Flux upper limit derivation}

In this appendix we show an alternative method to compute the non-detection likelihood presented in Section \ref{science}. The NITRATES analysis allows us to derive a flux upper limit at a given confidence level for each pixel of the GW sky localization, corresponding to the function $\phi_{\rm UL}( \rm RA, Dec)$ defined in Eq.~(\ref{phi_ul}). Then combined probability of being located in the pixel $x_i$ and to have a non-detectable EM emission is
\begin{equation}
\label{w_ul}
   P(\text{non-det},x_i) \propto P_{\text{GW}}( x_i) P[\phi<\phi_{\rm UL}( x_i)] \Delta \Omega_i,
\end{equation}
where
\begin{equation}
     P[\phi<\phi_{\rm UL}( x_i)] = (1-f) + f \int_0^{\phi_{\rm UL}(x_i)}P(\phi|x_i)\text{d}\phi,
\end{equation}
and $\Delta \Omega_i$ is the area of the pixel.
Here we express $P(\phi|x_i)$ as the conditional flux probability distribution, namely the flux probability distribution assuming that the GW source is contained in the pixel $x_i$. To compute the $P(\phi|x_i)$, for a fixed luminosity $L$, the luminosity distance is extracted from the the conditional probability distribution $P(D_{\rm L}|x_i)$, which is derived from the GW sky localization. Finally the overall non-detection probability is obtained integrating Eq.~(\ref{w_ul}) over the full sky:
\begin{equation}
\label{p_nondet}
     P(\text{non-det}|f,L_0) = \sum_{x_i} P_{\text{GW}}(x_i) P[\phi<\phi_{\rm UL}( x_i)]\Delta \Omega_i = (1-f) +f \Big [\varepsilon_{\Earth}+\sum_{x_i \notin \Omega_{\Earth}} P_{\text{GW}}(x_i) \Delta \Omega_i \int_0^{\phi_{\rm UL}(x_i)}P(\phi| x_i) \text{d}\phi\Big ],
\end{equation}
where we have used that
\begin{equation}
    \int_0^{\phi_{\rm UL}(x_i)}P(\phi| x_i) \text{d}\phi =1  \text{ if }x_i \in \Earth, \quad \text{  and  }\sum_{x_i \in \Omega_{\Earth}} P_{\text{GW}}(x_i) \Delta \Omega_i = \varepsilon_{\Earth}.
\end{equation}
The resulting probability of non-detecting any EM emission in correspondence to a GW trigger with a given $p_{\rm astro}=\pi_i$ is
\begin{equation}
\label{p_nodet1}
     P(\text{non-det}|f,L_0, \pi_i) = (1-\pi_i) + \pi_i P(\text{non-det}|f,L_0).
\end{equation}
Eq.~\eqref{p_nondet} has to be compared with the method used in Section \ref{science}, where instead we used the approximation:
\begin{equation}
\label{p_phi}
     P(\text{non-det}|f,L_0) = (1-f) +f \int_0^{\phi_{\rm UL}}P(\phi) \text{d}\phi,
\end{equation}
with 
\begin{equation}
     \phi_{\rm UL}= \int_{\Omega \notin \Omega_{\Earth}} \phi_{\text{UL}}(\Omega) P_{\text{GW}}(\Omega) \text{d} \Omega,
\end{equation}
and $P(\phi)$ is obtained extracting $D_{\text{L}}$ from the full sky marginalized distribution $P(D_{\text{L}})$, corresponding to
\begin{equation}
P(D_{\rm L}) = \sum_{x_i } P_{\text{GW}}(x_i) P(D_{\rm L}|x_i)\Delta \Omega_i.
\end{equation}
The two methods give comparable results in the assumption that the following approximation is valid:
\begin{equation}
\int_0^{\phi_{\rm UL}}P(\phi) \text{d}\phi \approx 
\varepsilon_{\Earth}+\sum_{x_i \notin \Omega_{\Earth}} P_{\text{GW}}(x_i) \Delta \Omega_i \int_0^{\phi_{\rm UL}(x_i)}P(\phi| x_i) \text{d}\phi.
\end{equation}
For completeness, we clarify here the main differences in the two methods. 
\subsection*{Method 1:}
This is the method used in Section \ref{science} and is based on the following steps:
\begin{enumerate}
    \item The marginalized upper limit $\phi_{\rm UL}$ is computed over the full sky, weighting by the GW sky localization.
    \item Once $L_0$ is fixed, the flux probability distribution $P(\phi)$ is computed extracting randomly $D_{\rm L}$ from the $P(D_{\rm L})$, the latter corresponding to the posterior distribution of the GW luminosity distance, marginalized over the full sky (excluding the part occulted by the Earth).
    \item The integral $\int_0^{\phi_{\rm UL}}P(\phi) \text{d}\phi$ which appears in Eq.~\eqref{p_phi} is evaluated counting the fraction of simulated events that have a predicted flux below the sky-averaged upper limit $\phi_{\rm UL}$.
\end{enumerate}
\subsection*{Method 2:}
This is the method presented in this appendix and summarized by Eqs.~\eqref{p_nondet} and \eqref{p_nodet1}, consisting in the following procedure:
\begin{enumerate}
    \item A set of sources is injected in space and the distribution follows the volumetric probability distribution of the GW candidate. First the coordinates of the injected source are extracted from the sky localization $P_{\text{GW}}(\rm RA, Dec)$, then for each position the distance is extracted according to the conditional probability $P(D_{\text{L}}|\rm RA, Dec)$.
    \item For each injected source, once the luminosity $L_0$ is fixed, the predicted flux is compared with the coordinates-\text{d}ependent BAT upper limit map $\phi_{\rm UL} (\rm RA,Dec)$.
    \item We define $\rho_{\notin \Earth}$ the fraction of all the sources injected which are not occulted by the Earth and also have a predicted flux below $\phi_{\rm UL} (\rm RA,Dec)$. Given this definition, we have:
    \begin{equation}
    \label{mah}
        \sum_{x_i \notin \Omega_{\Earth}} P_{\text{GW}}(x_i) \Delta \Omega_i \int_0^{\phi_{\rm UL}(x_i)}P(\phi| x_i) \text{d}\phi = (1-\varepsilon_{\Earth}) \rho_{\notin \Earth}.
    \end{equation}
\end{enumerate}
The last equality can be justified considering that, if for each pixel $i$ we inject $N_{\text{tot}, i}$ sources, we can define $\rho_i=N_{\text{ND},i}/N_{\rm \text{tot},i}$, where $N_{\text{ND},i}$ is the fraction of injected sources that are not detected, i.e., with a predicted flux below $\phi_{\rm UL}(x_i)$. Therefore:
    \begin{equation}
        \int_0^{\phi_{\rm UL}(x_i)}P(\phi| x_i) \text{d}\phi = \rho_i.
    \end{equation}
Let us call $N_{ \rm tot}$ the total number of sources injected on the full sky. Then we have
    \begin{equation}
        N_{\rm \text{tot},i} = N_{\rm \text{tot}} P_{\text{GW}}(x_i)\Delta \Omega_i,
    \end{equation}
and therefore
    \begin{equation}
        \sum_{x_i \notin \Omega_{\Earth}} P_{\text{GW}}(x_i) \Delta \Omega_i \int_0^{\phi_{\rm UL}(x_i)}P(\phi| x_i) \text{d}\phi = \sum_{x_i \notin \Omega_{\Earth}} \frac{N_{\text{tot},i}}{N_{\rm tot}}\rho_i=\dfrac{1 }{N_{\rm tot}}\sum_{x_i \notin \Omega_{\Earth}}N_{\text{ND},i}.
    \end{equation}
Since the total number of injected sources not occulted by Earth are $N_{\rm tot, \notin \Earth}=(1-\varepsilon_{\Earth})N_{\rm tot}$, and using that 
\begin{equation}
    \rho_{\notin \Earth}=\frac{1}{N_{\rm tot, \notin \Earth}}\sum_{x_i \notin \Omega_{\Earth}} N_{\text{ND},i},
\end{equation}
we finally recover Eq.~\eqref{mah}.

In order to quantify the difference between the two methods, the following test is performed. Having fixed the two parameters $f$ and $L_0$, we compute the likelihood $\mathcal{L}$ for the two methods and we derive the quantity
\begin{equation}
    \varepsilon_{\mathcal{L}} = 2\frac{\text{abs} (\mathcal{L}_1-\mathcal{L}_2)}{\mathcal{L}_1+\mathcal{L}_2}.
\end{equation}
Here we use the subscripts 1 and 2 for the respective methods. Both likelihoods are computed considering only BBH candidates with $p_{\rm astro}>0.5$. In both cases, the total number of injected sources for each BBH candidate is $N_{\rm tot}=1000$. The distribution of $\varepsilon_{\mathcal{L}} $ is evaluated sampling randomly $f$ and $L_0$, for a total of 100 sampled pairs ($f,L_0$). We obtain that the median value of $\varepsilon_{\mathcal{L}}$ is 0.04 and that in $\sim 80\%$ of the sampled cases $\varepsilon_{\mathcal{L}}<0.2$. Since the difference between the two methods is limited and since the Method 2 is more computationally expensive, all the results are used adopting Method 1.

\newpage

\clearpage

\LTcapwidth=\textwidth

\begin{longtable}{cccccccccc}
\caption*{List of 636 low latency GW triggers analyzed using NITRATES is shown along with their respective $p_{\rm astro}$ values and 15--350~keV band flux upper limits. The maximum $\sqrt{\text{TS}} $ is indicated for all the triggers with successful NITRATES results. Observations corresponding to triggers with insufficient exposure time during the BAT pointing mode do not have valid  NITRATES results or flux upper limits. For those triggers that do have NITRATES results but fail to meet the criterion for a full likelihood analysis, the max $\sqrt{\text{TS}}$ is indicated as NFL (No Final Likelihood).  The GW triggers from the \textit{Burst} pipeline do not have associated $p_{\rm astro}$ values and are therefore left blank. The fraction of the GW sky posterior distribution inside the BAT coded FOV and the fraction of the GW posterior occulted by the Earth are denoted by $\varepsilon_{\rm in \, BAT}$ and $\varepsilon_{\Earth}$, respectively.} \\

 \hline
 SID & Time  & GW FAR & group & $p_{\rm astro}$ & Class & $\sqrt{\text{TS}}$ & Flux UL & $\varepsilon_{\rm in \, BAT}$ & $\varepsilon_{\Earth}$  \\
    &(UTC)&(Hz)&&&&& (erg~cm$^{-2}$~s$^{-1}$)& (\%)& (\%)\\

     \hline
S190701ah & 2019-07-01T20:33:07 & 1.92$\times$10$^{-8}$ & CBC & 0.934 & BBH & 6.4 & 1.58$\times$10$^{-7}$ & 99.42 & 0
	\\
S190816i & 2019-08-16T13:04:31 & 1.44$\times$10$^{-8}$ & CBC & 0.833 & NSBH & - & - & - & -
	\\
S190828af & 2019-08-28T17:51:02 & 1.83$\times$10$^{-5}$ & CBC & 0.012 & BNS & - & - & - & -
	\\
S190829p & 2019-08-29T13:49:01 & 2.59$\times$10$^{-6}$ & CBC & 0.062 & Mass Gap & 5.8 & 2.05$\times$10$^{-7}$ & 58.03 & 0.94
	\\
S190830y & 2019-08-30T15:07:04 & 7.70$\times$10$^{-8}$ & CBC & 0.382 & Mass Gap & 5.7 & 1.45$\times$10$^{-7}$ & 78.14 & 5.94
	\\
\end{longtable}

\begin{table*}
    \centering
        \caption{Details of the O3 candidates confirmed by the offline analysis and with a $p_{\rm astro}>0.5$, for which GUANO data dumps are available. The reported $p_{\rm astro}$ and FAR are relative to the pipeline with the highest $p_{\rm astro}$. If two pipelines have equal $p_{\rm astro}$, we select the one with the highest SNR. The GW FAR, $p_{\rm astro}$ and Class details are quoted from \citet{GWTC-2.1PhysRevD.109.022001} and \citet{GWTC3-2023PhRvX}.}
    \begin{tabular}{lllccccr}
\hline
\hline
     SID   & GW name &FAR & Group & $p_{\rm astro}$ & Class &  $p_{\rm Class}$ & Pipeline \\
         &    &       (Hz) &    &       &         &   \\
\hline 
 S190701ah	& GW190701$\_$203306	& $ 1.79 \times 10^{-8} $ &	CBC & $>$0.99  & BBH	 		&  1.00   & PyCBC-BBH		\\
S190915ak  		 & GW190915$\_$235702 	&   $ 2.22\times 10^{-12} $	&	  CBC 	& 	$>$0.99 &	BBH & 1.00 & PyCBC-BBH   \\
S190930s	  	 & GW190930$\_$133541 	&   $ 3.81\times 10^{-10} $	&	  CBC 	& 	$>$0.99 &	BBH & 0.85 & PyCBC-BBH   \\
S191127p	  	 & GW191127$\_$050227 	&   $ 1.29\times 10^{-7} $	&	  CBC 	& 	0.74 &	BBH & 0.74 & PyCBC-BBH \\
S191204r	  	 & GW191204$\_$171526 	&   $ 1.86\times 10^{-13} $	&	  CBC 	& 	$>$0.99 &	BBH & 1.00 & MBTA 			\\
S191216ap  		 & GW191216$\_$213338 	&   $ 2.96\times 10^{-11} $	&	  CBC 	& 	$>$0.99 &	BBH & 1.00 & MBTA 			\\
S200128d	  	 & GW200128$\_$022011 	&   $ 1.36\times 10^{-10} $	&	  CBC 	& 	$>$0.99 &	BBH & 1.00 & PyCBC-BBH \\
S200129m	  	 & GW200129$\_$065458 	&   $ 9.03\times 10^{-41} $	&	  CBC 	& 	$>$0.99 &	BBH & 1.00 & GstLAL 		\\
S200208q	  	 & GW200208$\_$130117 	&   $ 9.84\times 10^{-12} $	&	  CBC 	& 	$>$0.99 &	BBH & 1.00 & PyCBC-BBH 	\\
S200216br  		 & GW200216$\_$220804 	&   $ 1.11\times 10^{-8} $	&	  CBC 	& 	0.77 &	BBH & 0.77 & GstLAL 		\\
S200220ad  		 & GW200220$\_$061928 	&   $ 2.16\times 10^{-7} $	&	  CBC 	& 	0.62 &	BBH & 0.62 & PyCBC-BBH 	\\
S200225q	  	 & GW200225$\_$060421 	&   $ 2.79\times 10^{-11} $	&	  CBC 	& 	$>$0.99 &	BBH & 1.00 & cWB 			\\
S200302c	  	 & GW200302$\_$015811 	&   $ 3.54\times 10^{-9} $	&	  CBC 	& 	0.91 &	BBH & 0.91 & GstLAL 		\\
S200322ab  		 & GW200322$\_$091133 	&   $ 1.44\times 10^{-5} $	&	  CBC 	& 	0.62 &	BBH & 0.62 & MBTA			\\	

        \hline
        \hline
    \end{tabular}

    \label{tab:gwtc3}
\end{table*}

\clearpage
\begin{longtable}{lllccrcr}
    \caption*{List of the O3 candidates confirmed by the offline analysis with $p_{\rm astro}<0.5$, for which GUANO data dumps were available. GW FAR, $p_{\rm astro}$, Class and $p_{\rm Class}$ are reported from \cite{GWTC3-2023PhRvX}. CBC or Burst group categories are quoted as per the offline analysis and not from the low-latency information.}\\
    
    \hline
         SID & Time & Group & GW FAR & $p_{\rm astro}$ & Class & $p_{\rm Class}$  & Pipeline \\
         & (UTC)&  & (Hz)&&&&\\
 \hline
  S190906ah	& 2019-09-06T20:05:00	& CBC	 & $ 5.66 \times 10^{-6} $ &	$ 2.38 \times 10^{-3} $	  & BBH	 		& $ 1.51 \times 10^{-3} $  & GstLAL		\\
S191106r	& 2019-11-06T18:41:51	& CBC	 & $ 1.18 \times 10^{-5} $ &	$ 1.97 \times 10^{-2} $	  & BBH	 		& $ 1.97 \times 10^{-2} $  & PyCBC-BBH		\\
S191116ac	& 2019-11-16T14:21:55	& CBC	 & $ 8.25 \times 10^{-6} $ &	$ 2.10 \times 10^{-4} $	  & NSBH	 	& $ 1.67 \times 10^{-4} $  & PyCBC-broad		\\
S191121bt	& 2019-11-21T16:45:42	& CBC	 & $ 5.44 \times 10^{-6} $ &	$ 4.13 \times 10^{-3} $	  & BBH	 		& $ 4.13 \times 10^{-3} $  & GstLAL		\\
S191208b	& 2019-12-08T02:02:15	& CBC	 & $ 2.10 \times 10^{-5} $ &	$ 4.26 \times 10^{-4} $	  & BNS	 		& $ 4.26 \times 10^{-4} $  & PyCBC-broad		\\
\end{longtable}

\newpage

\begin{table*}
    \centering
        \caption{Details of the joint FAR computed according to the procedure detailed in Section \ref{sec:far} for all the triggers with FAR$_{\rm GRB,\rm max}<10^{-3}$ Hz. The RAVEN alert is, by definition, evaluated considering only information received in low latency. The events marked with a (*) are GW candidates with $p_{\rm astro}>0.5$.}
    \begin{tabular}{llccrcc}
    \hline
    \hline
Name        & GW FAR    &  Group  &    Class    &   $\sqrt{\text{TS}}$    &   Joint FAR &  Raven Alert \\

        & (Hz)   &    &        &       &   (Hz) &   \\

\hline

S190919ag	&  $ 3.42 \times 10^{-6	} $ & Burst	&  -			& 7.23	& $ 2.16 \times 10^{-7	} $ & no	\\
S190919au	&  $ 3.24 \times 10^{-6	} $ & Burst	&  -			& 9.44	& $ 7.36 \times 10^{-9	} $ & no	\\
S190919u	&  $ 8.18 \times 10^{-6	} $ & Burst	&  -			& 8.00	& $ 8.56 \times 10^{-8	} $ & no	\\
S190930t	&  $ 1.54 \times 10^{-8	} $ & CBC	&  	NSBH		& 14.60	& $ 5.97 \times 10^{-11}  $ & yes	\\
S191110x	&  $ 2.93 \times 10^{-11} $ & CBC	&  	Mass Gap		& 7.17	& $ 9.59 \times 10^{-12}  $ & yes	\\
S191212l	&  $ 9.31 \times 10^{-6	} $ & CBC	&  	Mass Gap		& 7.16	& $ 4.62 \times 10^{-7	} $ & no	\\
\hline
\hline
    \end{tabular}

    \label{tab:jointfar}
\end{table*}

\end{document}